\documentclass{elsart}
\usepackage{amsmath,amssymb}
\usepackage{longtable}
\newif\iffigs
\figstrue

\def\pt{{\partial_t}}
\def\pdm{{\partial_x^{-4}}}
\def\pem{{\partial_x^{-5}}}
\def\pfm{{\partial_x^{-6}}}
\def\pcm{{\partial_x^{-3}}}
\def\pxxm{{\partial_x^{-2}}}
\def\pxm{{\partial_x^{-1}}}
\def\px{{\partial_x}}
\def\pxx{{\partial_x^2}}

\def\pd{{\partial_x^4}}

\def\lc{{\lambda}}
\def\f#1#2{{\frac{#1}{#2}}}

\def\pr#1{{\left( #1 \right)}}
\def\va{{v_a}}
\def\vb{{v_b}}
\def\va#1{{v_{a #1}}}
\def\vb#1{{v_{b #1}}}
\def\vfa{{\varphi_a}}
\def\vfb{{\varphi_b}}
\def\vpa#1{{\varphi_{a #1}}}
\def\vpb#1{{\varphi_{b #1}}}
\def\vvv{{\overline{v}}}
\def\vv#1{{\overline{v}^{(#1)}}}
\def\vur{{\overline{v_r}^{(1)}}}
\def\vub{{\overline{v_\beta}^{(1)}}}
\def\c#1{{c_{#1}}}
\def\Q#1{{Q^{(#1)}}}
\def\Gr#1{{\mathcal{G}_{#1}}}
\def\FGr#1{{\widehat{\mathcal{G}}_{#1}}}
\def\tv{{\tilde{v}}}
\def\Lr#1{{\mathcal{L}_{#1}}}

\iffigs
\usepackage{graphicx}
\graphicspath{{figs/}}
\fi

\begin{document}
\bibliographystyle{elsart-num}
\runauthor{Legras and Villone}
\begin{frontmatter}
\title{Dispersive and friction-induced stabilization of an inverse
cascade. The theory for the Kolmogorov flow in the slightly 
supercritical regime}
\author[LMD]{Bernard Legras}
\ead{legras@lmd.ens.fr}
\author[ICG]{Barbara Villone}
\address[LMD]{Laboratoire de M\'et\'eorologie Dynamique, UMR8539,
24 rue Lhomond, 75231 Paris Cedex 5, France}
\address[ICG]{Istituto Cosmogeofisica, CNR, Corso Fiume 4,
10133 Torino, Italy}
\date{22 May 2002}

\begin{abstract} 
  We discuss the stabilisation of the inverse cascade in the large
  scale instability of the Kolmogorov flow described by the complete
  Cahn-Hilliard equation with inclusion of $\beta$ effect, large-scale
  friction and deformation radius.  The friction and the $\beta$
  values halting the inverse cascade at the various possible
  intermediate states are calculated by means of singular
  perturbation techniques and compared to the values resulting from
  numerical simulation of the complete Cahn-Hilliard equation. The
  excellent agreement validates the theory. Our main result is that 
  the critical values of friction or $\beta$ halting the inverse cascade
  scale exponentially as a function of the jet separation in the final
  flow, contrary to previous theories and phenomenological approach.

\end{abstract}
\end{frontmatter}

\section{Introduction}
\label{s:intro}

Inverse cascades are a common feature in the large-scale velocity and
magnetic fields of geophysical, planetary and astrophysical
two-dimensional flows.  Their halting by spontaneous formation of
zonal jets has been object of great interest and considerable work by
many scientists, see among other
Refs~\cite{Rhines:75,Williamson:78,Rhines:94,Vallis:93,Manfroi:99}.
The phenomenon of an halted inverse cascade could play a role in the 
atmosphere of Jupiter and other Jovian
planets which exhibit jet streams of east-west and west-east
circulation.

Frisch et al. \cite{Frisch:96} showed that the inverse cascade of the
large-scale nonlinear instability of the Kolmogorov flow described by
the Cahn--Hilliard equation
\cite{Meshalkin:61,Sivashinsky:85,Nepomnyashchy:76} may be stopped by
the dispersive Rossby waves, i.e. by the so-called the $\beta$-effect. We
recall here that in the absence of any stabilizing effect the inverse
cascade proceeds by visiting a family of metastable states with
increasing scale until the final largest scale is reached
\cite{She:87}.  In a later paper, Legras et al. \cite{Legras:99} have
used singular perturbation techniques to calculate the range of the
$\beta$ values stopping the inverse cascade in one of the otherwise
(if $\beta$ = 0) metastable state.  Their result was different from
that obtained by the standard phenomenology based on dimensional
arguments \cite{Rhines:75} which fails because it does not take into
account the strong suppression of non linearities in the metastable
states of the inverse cascade.  

In Refs.~\cite{Frisch:96,Legras:99} the forcing maintaining the basic
flow was chosen parallel to the planetary vorticity contours and there
was no friction or advection. In a more realistic setup, Manfroi and
Young \cite{Manfroi:99} studied the stability of a forced meridian
flow on a $\beta$-plane pushing the fluid across the planetary
vorticity contours, and including both friction and advection by a
mean flow.  They obtained a complete amplitude equation for the
leading order perturbation, from which, with some formal modifications
and with a slightly different interpretation of the parameters, the
amplitude equation of Ref.~\cite{Frisch:96} can be recovered (see
Section.\ref{s:basiceq}).  Using this equation, but disregarding the
dispersive contribution from $\beta$-effect, they showed that random
initial perturbations rapidly reorganize into a set of fast and narrow
eastward jets separated by slower and broader westward jets, followed
then by a much slower adjustment of the jets, involving gradual
migration and merger. The stabilization discussed in
Ref.~\cite{Manfroi:99} is only due to friction and not to dispersive
effects.

In this paper we present a systematic approach to the study of
stabilization of the inverse cascade in the supercritical regime of
the large-scale Kolmogorov flow provided both by friction and $\beta$
effect.  The stabilization by $\beta$ effect already discussed in
\cite{Legras:99} will be here presented in a greater detail and
stabilization by friction will be discussed in the same mathematical
approach as for the $\beta$ case.  The mathematical framework is based
on the kink dynamics introduced by Kawasaki and Ohta
\cite{Kawasaki:82} to describe the solutions to the Cahn--Hilliard
equation; singular perturbation technique is used to study the
stability of these solutions with respect to small perturbations due
to friction and $\beta$ effect. The perturbative calculations are
performed analytically for large wavenumbers and numerically for all
cases. The results are compared together and with direct numerical
stability and time-dependent solutions of the amplitude equation.

The paper is organized as follows: in Section~\ref{s:compCH} the
Cahn--Hilliard equation in its {\em complete} form is presented; by
complete in this context we mean that $\beta$, friction, advection
velocity and deformation terms have been added to the standard
Cahn--Hilliard equation.  The perturbation to the steady metastable
solution of the Cahn--Hilliard equation by small $\beta$ and a
friction terms is presented in Section~\ref{s:persta}.
Section~\ref{s:stab} discusses the stability of the steady metastable
solutions and provides the main results of this work.
Section~\ref{s:numerical} presents the techniques used to solve
numerically the perturbation and the time-dependent problem, and
compares the results of those calculations and the analytical results.
Section~\ref{s:conclu} offers a summary and conclusions.

We do not give the detailed presentation of the mean advection effect,
which introduces considerable additional algebraic complications, in
order to focus here on the essential mechanisms that stabilizes the
Cahn-Hilliard cascade. The main results in presence of mean advection
are, however, given without demonstration in Appendix~\ref{s:meanadv}.

\section{The complete Cahn--Hilliard equation}
\label{s:compCH}

\subsection{Basic equation}
\label{s:basiceq}

Our starting point is the large-scale Kolmogorov flow in its slightly
supercritical regime, which is described by the Cahn-Hilliard equation
\cite{Meshalkin:61,Sivashinsky:85,Nepomnyashchy:76}.  We recall here
that the basic Kolmogorov flow, ${\bf u} = (\cos y,0)$, is maintained
by a force ${\bf f}= \nu(\cos y,0)$ against viscous dissipation.  This
flow exhibits a large-scale instability of the negative eddy viscosity
type when the kinematic viscosity $\nu$ is slightly below the critical
value $\nu_c=1/\sqrt2$. 

Taking further into account $\beta$ effect, friction $r$, external
deformation radius $1/S$ \footnote{The external deformation radius
  accounts for the inertial large-scale effect of a free surface in
  geophysical flows \cite{Pedlosky:87}.} and advection effect due to
a non zero mean velocity $\gamma$ , we obtain by multi-scale techniques
the following adimensional equation for the leading order large-scale
perturbation :
\begin{equation}
   \pt (1 - S^2 \pxxm) (v-\gamma) = \lc \pxx W'(v) - \lc \pd v 
   -\beta \pxm (v-\gamma) - r v,
   \label{vbetach}
\end{equation}
where $\pxm$ denotes the integration in $x$ defined for the family of
functions with zero average over the interval $[0,L]$.  The constants
in (\ref{vbetach}) are
\begin{equation}
   s=\frac{1}{\sqrt{3}},\qquad
   \Gamma=\sqrt{\frac{3}{2}},\qquad 
   \lambda_3=\frac{3}{\sqrt2}.
   \label{deflam123}
\end{equation}
 and the potential $W(v)$ is
\begin{equation} 
   W(v)= \frac{s^2}{2 \Gamma^2} v^{4} - s^2 v^2 
\label{Wdef} 
\end{equation}
Equation (\ref{vbetach}) was derived in \cite{Frisch:96} with $S=r=\gamma=0$.
In this derivation, the Kolmogorov basic flow is oriented in the zonal 
direction on the $\beta$-plane and therefore the large amplitude flow 
develops in the meridional direction. Using a more realistic setting
where the Kolmogorov basic flow is oriented in the meridional 
direction as an idealized baroclinic perturbation, and introducing friction 
and  mean advection velocity, 
Manfroi and Young~\cite{Manfroi:99} derived the following amplitude equation
\begin{multline}
   \partial_\tau A = - r A - (2-{\gamma^*}^2) 
   \partial_{\eta^2} A - 3 \partial_{\eta^4} A \\ 
   + 2 \gamma^* \partial_\eta(\partial_\eta A^2)
   + \frac{2}{3} \partial_\eta(\partial_\eta A^3) 
   - \beta \partial_{\eta^{-1}} A 
   \label{eq:mf}
\end{multline}
where $\tau$ and $\eta$ are the temporal and spatial variables.  
In their study, the last term on the right-hand side, which is
the only dispersicve term in the equation, was set to zero.
 \footnote{The coefficient $\beta$ in the last term of the right hand side
  of (\ref{eq:mf}) is the product of the planetary vorticity gradient
  per the small angle between the Kolmogorov flow direction and the
  planetary vorticity gradient.  Therefore it can be either positive
  or negative unlike in the derivation of Ref.~\cite{Frisch:96} where it is
  always positive}.  By the change of variable $\px A=w-\gamma$ and
the rescaling $\tau= a t$, $\eta=bx$, $w=cv$ and $\gamma^*=c \gamma$
where $a=12 s^4 \lc / (2+{\gamma^*}^2)^2$,
$b=(6s^2/(2+{\gamma^*}^2))^{1/2}$ and $c=(3(2+{\gamma^*}^2)/(2
\Gamma^2))^{1/2}$, this equation is easily transformed into
(\ref{vbetach}) up to the term in $S$ which is only a trivial
modification \cite{Pedlosky:87}.  We see that the quadratic term in
(\ref{eq:mf}) disappears in this transformation.  The $\gamma$ term
represents the net advection velocity (see \cite{Manfroi:99}) and is also
the average value of $v$.

In a recent work \cite{Manfroi:02}, it is claimed that the Kolmogorov 
instability exhibits a singular limit when $\beta \leftarrow O$. This result
is, however, established for a very special situation which does 
not arise here \cite{Legras:02a}. 

\subsection{Kinks and antikinks}

The pure Cahn--Hilliard equation, is recovered from (\ref{vbetach}) by
setting $\beta=r=S=\gamma=0$. It admits a Lyapunov functional and is
therefore integrable \cite{Sivashinsky:85}.  
This property is preserved in the presence of
friction but is lost in the presence of $\beta$ 
\footnote{A Lyapunov formulation for the approximated equation is
  recovered, however, in the limit of large $\beta$ \cite{Frisch:96}.}. 
The solutions to the
pure Cahn--Hilliard equation live essentially, albeit some initial
transients, within a slow manifold of soliton-like solutions with an
alternation of plateaus $v=\pm \Gamma$, separated by alternating
positive and negative kinks, that we will call respectively kinks and
antikinks \cite{Bates:95} in the following. For large enough
separation between adjacent kinks, the kink centered in $x=x_j$ is
locally given by
\begin{equation}
   M_j(x)= \epsilon_j M(x-x_j)=\epsilon_j \Gamma \tanh s (x-x_j) \,,
   \label{kink}
\end{equation}
where 
$\epsilon _j = 1$ for a kink and $\epsilon _j = -1$ 
for an antikink \cite{Kawasaki:82}.
This solution satisfies the equation
\[
   -\pxx M_j + W'(M_j) = 0 \, .
\]

Within a $x$ periodic domain of period $L$, the Cahn--Hilliard equation 
exhibits
stationary metastable solutions of period $\Lambda=L/N$ with $N$ pairs
of alternating and equally spaced kinks and antikinks.  These fixed
points are unstable saddle points of the Lyapunov functional, except
for $N=1$ which corresponds to an absolute stable minimum.  The
temporal evolution characterized by a growing total energy is a
cascade of annihilations of kink-antikink pairs 
(see Fig.~\protect\ref{f:ANKINK}) leading
eventually to the gravest mode $N=1$ \cite{Kawasaki:82}.
It is shown in Appendix~\ref{aCH} that the local solution 
(\ref{kink}) is modified by terms of order $\exp(-s\Lambda)$ when
the periodicity is taken into account.

\begin{figure}
  \iffigs
    \centering
    \includegraphics*[width=10cm]{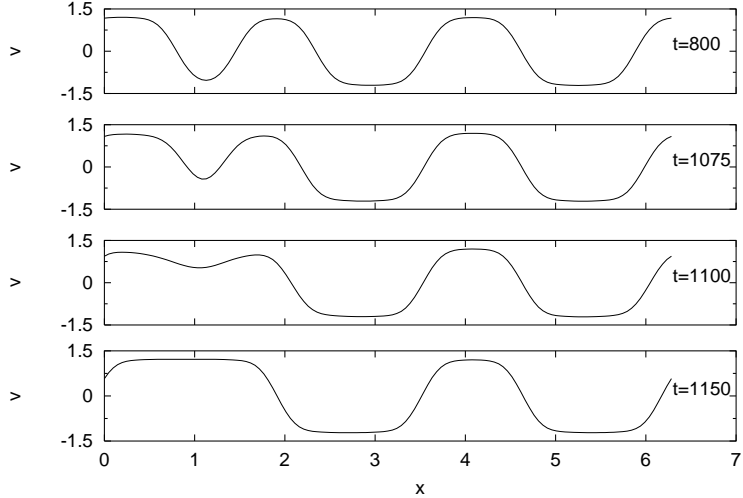}
  \else
    \drawing 100 10 {Kink-antikink annihilation}
  \fi
  \caption {Kink-antikink annihilation in a numerical simulation
  of the pure Cahn-Hilliard equation with $L$=76.95 .}
  \protect\label{f:ANKINK}
\end{figure}

\section{Perturbation of stationary solution under the
action of $\beta$ and friction}
\label{s:persta}

In order to study the modification of the stationary solutions to the
pure Cahn-Hilliard equation under the effect of a small $\beta$ or friction,
we multiply both terms by a small control parameter, $\varepsilon$, 
keeping $\beta$ and $r$ as O(1).

We put $\gamma = 0$, leaving the case $\gamma \ne 0$, which is 
technically much more involved, for Appendix~\ref{s:meanadv}.
We only notice here that $\gamma > 0$ breaks one symmetry and generates 
narrow intense westerlies and broad narrow easterlies \cite{Manfroi:99}.
  
It is convenient to integrate (\ref{vbetach}) twice in $x$ to obtain
\begin{equation}
   - \frac{1}{\lc} \pxxm (1 - S^2 \pxxm)
   \pt v - \varepsilon \frac{\beta}{\lc} \pcm v 
   - \varepsilon \frac{r}{\lc} \pxxm  v = \pxx v - W'(v) + h(t) \,,
   \label{bCH} 
\end{equation}
where $h(t)$ arises from integration in $x$.  Then the perturbed
solution is defined as $\vvv=\vv0 +\varepsilon \vv1 +
\varepsilon^2\vv2 + O(\varepsilon^3)$, where $\vv0$ satisfies the
stationary CH equation $-\pxx \vv0 + U'(\vv0)=0$.  We will distinguish
two cases for $S$: (i) $S=0$, associated with synoptic and subsynoptic
dynamics, and (ii) $S=O(1)$, associated with planetary motion (cf.
\cite{Pedlosky:87}).  When $\beta \ne 0$, we introduce the phase velocity
$c = \varepsilon \c1 + \varepsilon^2 \c2 + O(\varepsilon^{3})$ of the
traveling framework in which $\vvv$ is stationary.

\subsection{Order 1 perturbation}
\medskip

We first treat the case $S=0$.
The first order perturbation $\vv1$ satisfies the linear equation 
\begin{equation}
   \mathcal{F}\pr{\vv1} = \Q0      
   \label{baseper}
\end{equation}
with
\begin{align*}
   \mathcal{F}\pr{g}  &= \pxx g - W''_{0} g
   \, , \\
   W''_{0} &= W'' (\vv0) \, , \\
   \Q0 &= \frac{1}{\lc} (\c1 \pxm \vv0 - 
   \beta \pcm \vv0 - r \pxxm \vv0 ) 
   \, .
\end{align*}
 
$\px \vv0$ belongs to the kernel 
of $\mathcal{F}$: this can be easily verified 
by multiplying (\ref{baseper}) by $\px \vv0$ and integrating within 
the domain. 
After integration over the spatial period,
the  solvability condition for (\ref{baseper}) gives the first order 
contribution to the phase velocity
\begin{equation}
   \c1 = - \beta \frac{\int_0^L \pr{\pxm \vv0}^2 dx}
   {\int_0^L \pr{\vv0}^2 dx} \, .
   \label{speed} 
\end{equation}

The symmetry of the stationary Cahn--Hilliard equation with respect to $x$
reversal is inherited by $\vv0$. The solution is
antisymmetric with respect to kinks locations and symmetric with
respect to the middles of the plateaus. The perturbation $\vv1$ is
the sum of two parts
\[   \vv1 = \beta \vub + r \vur  \,, \]
where $\vur$ has the same symmetry as $\vv0$ and $\vub$ has opposite symmetry. 

It is shown in Appendix~\ref{aCH} that, up to errors of
$O(e^{-s\Lambda/2})$ the basic solution $\vv0$ can be approximated by a
series of jumps locally described by (\ref{kink}). Over each interval
$[x_j-\Lambda/4,x_j+\Lambda/4]$ centered on a kink in $x_j$ (see
(\ref{kink})), we have
\[ 
   \pxm \vv0 =  \epsilon_j \Gamma 
   \ln \pr{\frac{\cosh s x}{\cosh \f14 s \Lambda}} + O(e^{-s \Lambda/2})\,, 
\]
from which the velocity $\c1$ is readily calculated using (\ref{speed}).
After some algebra, we get
\begin{equation}
   \c1 = - \beta \pr{\frac{\Lambda^2}{48} - \frac{\pi^2}{12 s^2} +
   \frac{A + \f32 \zeta(3)}{\Lambda s^3}}\pr{
   1 - \frac{4}{\Lambda s}}^{-1} + O(e^{-s \Lambda/2}) \,,
   \label{c1}
\end{equation}
where $\zeta$ is the Riemann zeta function and $A=\int_0^\infty
\ln^2 (1 + e^{2 x}) dx = 0.150257 \cdots$.  The phase velocity is
directed to the left when $\beta>0$ and increases with the wavelength
$\Lambda$. The fact that the error in (\ref{c1}) is exponentially
small makes this expression very accurate even for not so large values
of $\Lambda$ as we shall check below.  By straightforward algebra, one
gets 
\begin{multline} 
  \Q0 = - \frac{\beta \Gamma}{2 \lc} |x| \pr{\frac{x^2}{3}
  -\f14 \Lambda |x| + \frac{1}{24} \Lambda^{2}} \\
  - \frac{r \Gamma}{2 \lc} 
  |x| \pr{ |x| - \f12 \Lambda } + O(\beta \Lambda^2 + r \Lambda)
  \,.  
\end{multline}
At distance from the kinks the ratio between the derivative and the
potential term in $\mathcal{F}$ is $O(1/\Lambda^2)$.  At leading order
in $1/\Lambda$:
\begin{align}
   \vub &= \frac{\Gamma}{24 s^2 \lc} |x| \pr{|x| - \f12 \Lambda}
   \pr{|x| - \f14 \Lambda} \, ,  \label{vub} \\
   \vur &= \frac{\Gamma}{8 s^2 \lc} |x| \pr{|x| - \f12 \Lambda}
   \, . \label{vur}
\end{align} 
These expressions are valid at $O(\Lambda)$ distance from the kinks.
It can be shown that at $O(1)$ distance from the kinks, $\vub$ is
$O(\Lambda^2)$, while $\vur$ is $O(\Lambda)$.

For $S=O(1)$, the leading contribution of friction is
unchanged, but the effect of $\beta$ is deeply affected. $\Q0$ is
modified as
\[
   Q^{(0)}_S = \frac{1}{\lc} \left(\c1 \pxm \vv0 - 
   (\beta + \c1 S^2) \pcm \vv0 - r \pxxm \vv0 \right) \, .
\]
Therefore, the phase speed is now 
\[
   \c1_S = - \frac{\beta \c1}{S^2 \c1 + \beta},
\]
where $\c1$ is given by (\ref{c1}), that is 
\begin{equation}
  \c1_S =  -\frac{\beta}{S^{2}} + \frac{48 \beta}{\Lambda^2 S^4} + 
  O\pr{\frac{\beta}{\Lambda^{3}}} \, .
  \label{c1S}
\end{equation}
Unlike the infinite radius case, the phase speed varies weakly 
with $\Lambda$. Reporting (\ref{c1S}) in $Q^{(0)}_S$ 
and solving for $\vub$, we obtain
\begin{equation}
   \vub = \frac{\Gamma}{s^2 S^2 \Lambda \lc} |x| \pr{|x| - \f14 \Lambda}
   \pr{1 - 2 \frac{|x|}{\Lambda}} + O(1)
   \label{vubS}
\end{equation}
at distance from the kinks. The correction to the stationary solution
scales as $\Lambda$ and is considerably reduced with respect to the 
case $S=0$, for which it scales as $\Lambda^3$.

\section{Stability}
\label{s:stab}

\subsection{Stability of the Cahn--Hilliard equation}
\label{s:CHstab}

The stability of the solution $v=0$ is easily obtained by linearizing
(\ref{vbetach}) in the Fourier domain.  When $r=0$, the solution is
unstable to all Fourier modes with wavenumber $0<k<k_m=\sqrt{2} s =
\sqrt{2/3}$. This result does not depend on the values of $\beta$ and
$S$. When $r>0$ the modes near $k=0$ and $k_m$ are stabilized and the
instability band in $k$-space shrinks as $r$ grows. The solution $v=0$
is stable for $r> r_0 = \lc s^4 = (3 \sqrt{2})^{-1}$.  The vicinity of
this value has been studied in \cite{Manfroi:99}.  We are here
interested in the limit of small $r$.

The stability of the non-zero stationary solutions to the Cahn--Hilliard
equation can be studied using the equation for kink motion derived in
Appendix~\ref{s:CH}. For an arbitrary perturbation, fast transients
dissipate rapidly, leaving only after a short time the part of the
perturbation that projects onto kink displacement.  The $j$th kink
being displaced by $\delta x_j$, the perturbation to $\vv0$ is
\begin{equation} 
   \delta v = - \sum_{\ell=0}^{2N-1} \px
   M_{\ell}\, \delta x_{\ell} \, .
   \label{deltav}
\end{equation}

Then using (\ref{motok}), the equation for the displacements is 
\begin{multline}
    -\frac{4 \Gamma^2}{\lc} \sum_{\ell=0}^{2N-1}
    \pr{(-1)^{j-\ell}\Gr2(x_j-x_{\ell})+ (-1)^{j-\ell} \frac{\pi^2}{12 L s^2}
    - \frac{1}{2 s}\delta_{j-\ell}} \delta \dot{x}_{\ell} \\
    = 64 s^3 \Gamma^2 e^{-s \Lambda} [ 2 \delta x_j - \delta x_{j+1}
    - \delta x_{j-1}] - 2 \Gamma \epsilon_j \delta h \,.
    \label{eq:pertu}
\end{multline}

It is convenient to use Fourier components defined as
\[ \delta x_j = \sum_{m=0}^{2N-1} \psi_m e^{i \pi\frac{mj}{N}} \,,\]
with $m \in [0,2N-1]$ and $\psi_{2N-m}= \overline{\psi_m}$.  The
equation for $\psi_m$ is obtained by multiplying (\ref{eq:pertu}) by
$\frac{1}{2N} e^{-i \pi\frac{mj}{N}}$ and summing over the $j$ from
$0$ to $2N-1$. Taking into account the regular alternation of kinks and
antikinks separated by intervals of length $L/2N$ in the basic
solution and using the Fourier transform of the Green function given
by (\ref{sg}), one obtains after some algebra:
\begin{multline}
   \label{eq:psim}
   \frac{1}{\lc}\pr{\frac{\Lambda}{1+\cos \theta_m} - \frac{2}{s}}
   \dot{\psi}_m
   = 128 s^3  e^{-s \Lambda} (1 - \cos \theta_m) \psi_m \\
   - 4\frac{N}{\Gamma} \delta h \; \delta(N-m) \, ,
\end{multline}
with $\theta_{m}= \pi m /N$.  The leading order form of
(\ref{eq:psim}) was given by Kawasaki and Ohta~\cite{Kawasaki:82} with a
factor 2 error (see Appendix \ref{s:CH}).

The first term in the right hand side of (\ref{eq:psim}) is
destabilizing the stationary solution with the eigenvalue
\begin{equation}
   \sigma_{0} = \frac{128 s^{3} \lc e^{-s \Lambda}}{\Lambda} \sin^2 
   \theta_m \pr{1 - \frac{2(1 + \cos \theta_m)}{s \Lambda}}^{-1} \, .
   \label{s0}
\end{equation}
This instability is responsible of the inverse cascade in the 
CH equation. Each value of $m$ is associated with a real eigenvalue of 
$\mathcal{F}$ and a dimension 2 eigenspace. It turns out that an appropriate 
basis of this eigenspace is provided by the couple of orthogonal vectors
\begin{align}
   \va (x) &= \sum_{j=0}^{2N-1} (-1)^j \cos j \theta_m \px M 
   (x-x_{j})  \, ,
   \label{defva} \\
   \vb (x) &= \sum_{j=0}^{2N-1} (-1)^j \sin j \theta_m \px M 
   (x-x_{j}) \, ,
   \label{defvb}
\end{align}
which are here given up to an error $O(e^{-s \Lambda/2})$. 
These expressions agree very well with the numerical solution shown in
figure~\ref{fig:vavb}.

\begin{figure}
  \iffigs
     \centering
     \includegraphics*[angle=-90,width=6cm]{statiog-4.1-a.epsi}
     \hspace{1.cm}
     \includegraphics*[angle=-90,width=6cm]{statiog-4.1-b.epsi}
  \else
     \drawing 100 10 {Eigenvectors va and vb}
  \fi
  \caption{Numerical calculation of $v_a$ and $v_b$ by discretization of 
  the eigenvalue problem for $\mathcal{F}$ (cf Section
  \ref{s:numpert}) with 128 points for $L=76.953$ and $m=4$.  (a) thin
  solid: $\vv0$; solid: $v_a$, dash: $v_b$; (b) thin solid: $\px
  \vv0$; solid: $v_a/\px \vv0$, dash: $v_b/\px \vv0$.  The scale is
  arbitrary for $v_a$ and $v_b$.  Even if the plateaus in $\vv0$ are
  very short for the chosen values of $L$ and $m$, the ratios in (b)
  show the staircase structure of $v_a$ and $v_b$ over the kinks as
  given in \protect{(\ref{defva}, \ref{defvb})} except where $\px
  \vv0$ vanishes in the middle of the plateaus.
  \label{fig:vavb}}
\end{figure}

The $\delta h$ contribution vanishes but on the mode $m=N$. The 
solution $v(x,t)$ must average to zero within the periodic interval
$[0,L]$. In terms of kink motion, this imposes the constraint
\[ \sum_{\ell=0}^{2N-1} (-1)^{\ell} \delta \dot{x}_{\ell} = 0 \,, \]
and thus, $\dot{\psi}_{N} = 0$.
The presence of $\delta h(t)$ in (\ref{eq:psim}) is required to impose 
this condition. 

\subsection{Stability of $\beta$-CH equation with infinite radius of 
deformation}

\subsubsection{Formulation of the problem}
The equation governing the perturbation $\delta v$ to $\vv0$ can be 
written conveniently for $\varphi = \pxm \delta v$
\begin{equation}
   \pt \varphi = \mathcal{L} \varphi \, ,
   \label{eq:stab}
\end{equation}
with 
\begin{equation}
   \mathcal{L}  =   - \lc \px (\pxx - W''(\overline{v})) \px 
   - c \px - \varepsilon \beta \pxm  - \varepsilon r \, .
   \label{L:def}
\end{equation}
The reason of using $\varphi$ instead of $\delta v$ is that $\mathcal{L}$ is 
auto-adjoint while the corresponding operator for $\delta v$ is not.

Unlike the case $\beta = 0$, the slow component perturbation to the
stationary solution of the $\beta$-CH equation does not reduce to the
simple motion of kinks. One has also to take into account the
dispersive effect of the $\beta$ term modifying the shape of the slow
modes and contributing to the stability.  Therefore, we expand
$\mathcal{L}$ as
\begin{equation}
   \mathcal{L} = \Lr0 + \varepsilon \Lr1 + \varepsilon^2 \Lr2 
   + O(\varepsilon^3) 
    \, , 
\end{equation}
with
\begin{align}
   \Lr0 &=  - \lc \px ( \pxx - U''_{0}) \px \, , 
   \label{L0:def} \\
   \Lr1 &=  \lc \px (W'''_{0} \vv1 \px) + \c1 \px - \beta \pxm - r
   \, , \label{L1:def}  \\
   \Lr2 &=  \lc \px (W'''_{0} \vv2 + \f12 W^{IV}_{0} \vv1^2 ) \px + 
   \c2 \px \, .
   \label{L2:def}
\end{align}

The eigenvalues of (\ref{eq:stab}) are perturbations of the 
eigenvalues of (\ref{eq:pertu}). For a given $m \neq N$, we obtain
\begin{equation}
   \sigma = \sigma_{0} + \varepsilon \sigma _{1} + i \varepsilon 
   \mu_{1} + \varepsilon \sigma_{2} + i \varepsilon \mu_{2} 
   + O(\varepsilon^{3}) \, .
\end{equation}
The functions $\vfa = \pxm \va{} $ and $\vfb = \pxm \vb{} $ are
orthogonal eigenmodes of $\mathcal{L}_0$ and, it turns out, an
appropriate Jordan basis for the perturbation problem. They are
respectively modified as $\vfa + \varepsilon \vpa1 + \varepsilon^2
\vpa2 + O(\varepsilon^3)$ and $\vfb + \varepsilon \vpb1 +
\varepsilon^2 \vpb2 + O(\varepsilon^3)$. The hierarchy of linear
problems is
\begin{align}
   \Lr0 \vfa & = \sigma_0 \vfa \, ,
   \label{eq:L0}  \\
   \Lr0 \vpa1 + \Lr1 \vfa &=  \sigma_0 \vpa1 + \sigma_1 
   \vpa - \mu_{1} \vfb \, ,
   \label{eq:L1}  \\
   \Lr0 \vpa2+ \Lr1 \vpa1 + \Lr2 \vfa &= \sigma_0 \vpa2
   + \sigma_{1} \vpa1 + \sigma_{2} \vfa - \mu_{1} \vpb1 - \mu_2     
   \vfb \, .
   \label{eq:L2}
\end{align}
and similar equations for $\vpb{i}$. 

\subsubsection{Stabilization by friction at first order}

The first order corrections of the eigenvalue $\sigma$ are obtained as 
solvability conditions of (\ref{eq:L1}) by
\begin{align}
   <\vfa,\Lr1 \vfa> &=  \sigma_{1}  <\vfa,\vfa> \, ,
   \label{eq:s1}  \\
   <\vfb,\Lr1 \vfa> &= - \mu_{1}  <\vfb,\vfb> \, ,
   \label{eq:m1}
\end{align}
where $<f,g> \equiv \frac{1}{L}\int_{0}^{L} f(x) g(x) dx$.
In $<\vfa,\Lr1 \vfa>$, the contributions from $\beta$ immediately 
vanish by integration. We have, after integration by part
\begin{equation} 
   <\vfa,\Lr1 \vfa> =  -\lc < \va{}^2, W'''_0 \vv1 > 
   + r <\va{}, \pxxm \va{}> 
   \, . \label{paL1pa}
\end{equation}
The first contribution to the right hand side of (\ref{paL1pa}) can be 
reduced to an integral over a single kink by summing the trigonometric 
factors arising from $\va{}^2$:
\begin{equation} 
   < \va{}^2, W'''_{0} \vv1 > = \frac{r}{\Lambda} \int W'''_{0} 
   \vur (\px M_{j})^2 dx \, .
\end{equation}
Here $j$ labels an arbitrary kink and the integral bounds do not need 
to be specified since $\px M_{j}$ decays exponentially on both side. 
This contribution is further transformed using
\[
   \int W'''_{0} \vur (\px M_j)^2 dx =   
   \frac{1}{\lc} \int \pxm \vv0 \px  M_j dx \, ,
\]
which is valid up to exponentially small errors. Finally the 
contribution is reduced to a non local integral by part,
\[
   \frac{1}{\Lambda} \int \pxm \vv0 \px  M_j dx = \frac{1}{2 \Lambda} 
   \int_{x_{j}- \f12 \Lambda}^{x_{j}+ \f12 \Lambda} \px \vv0 \pxm \vv0 
   dx =  -\f12 \Gamma^2 + \frac{2 \Gamma^{2}}{\Lambda s}\, . 
\]
Combining this with (\ref{vanva}) and (\ref{sg}), we obtain
\begin{multline} 
  \sigma_{1} = -r \sin^2 \f12 \theta_{m} \\
  + \frac{4r}{\Lambda s} \left(1 + \cos^{2} \f12 \theta_{m}\right)
  \left(1 - \frac{4}{\Lambda s } \cos^{2} \f12 \theta_{m}\right)^{-1}
  + O(e^{-s \Lambda/2}) \, .
   \label{s1}
\end{multline}

In a similar way, we have
\begin{multline}
   <\vfb,\Lr1 \vfa> = - \lc <\va{} \vb{} , W'''_0 \vv1> \\
   + \c1  <\va{},\pxm \vb{}> - \beta <\va{} , \pcm \vb{}> \, .
   \label{pbL1pa}
\end{multline}
The first term in the right hand side of (\ref{pbL1pa}) vanishes after
the trigonometric summation. Using (\ref{vanva}-\ref{vanvb}) and
(\ref{sf}-\ref{sh}), we obtain
\begin{multline}
  \mu_1 = \left(-\frac{2 \c1 \Gamma^2}{\Lambda} t - 
  \frac{\Gamma^2 \Lambda}{8} t(1+t^2)
  + \frac{\pi^2 \Gamma^2}{6 s^2 \Lambda} t \right)\\
  \times \left( \frac{\Gamma^2}{2} (1+t^2) - \frac{2 \Gamma^2}
  {s\Lambda } \right)^{-1} + O(e^{-s \Lambda/2}) \, ,
  \label{m1a} 
\end{multline}
or, retaining only the first three orders of the 
expansion,
\begin{equation}
   \mu_{1} = \beta \pr{- \frac{\Lambda t (2 + 3 t^2)}{12 (1+ t^2)} 
   - \frac{t (1+2t^{2})}{3s(1+t^{2})^{2}}+ \frac{4}{3\Lambda s^{2}}
   \frac{t^{5}}{(1+t^{2})3}} +  O \pr{\frac{\beta}{\Lambda^{2}}}\, ,
   \label{m1b}
\end{equation}
with $t= \tan \pi m/ 2 N$. 

It is interesting to notice that the nonlinear contribution $<
\va{}^2, U'''_{0} \vv1 >$ is destabilizing the stationary solution.
However, the direct linear damping by friction remains larger and the
total effect of friction is always stabilizing. Therefore, the
$m$-mode perturbation to the stationary solution is stabilized by
friction for
\begin{equation} 
   r > r_{c} = 512 \frac{e^{-s \Lambda}}{\Lambda} s^3
   \lc \cos^2 \frac{\pi m}{2 N} \, ,
   \label{critr} 
\end{equation}
at leading order.

The corresponding result when $\gamma \neq 0$ is given in 
(\ref{eq:rcadv}). The mean advection decreases the value
of friction necessary to stabilize a given wavenumber.

\subsubsection{Stabilization by $\beta$-effect at second order}

At second order in $\varepsilon$, $\sigma_{2}$ is solution of the 
solvability condition. Here we set $r$ to zero  for simplification
as stabilization by $r$ is already obtained at first order.
\begin{equation}
   <\vfa, \Lr1 \vpa1 > + < \vfa, \Lr2 \vfa >\; = \sigma_{2} <\vfa, 
   \vfa > - \mu_{1} <\vfa,\vpb1>  \, .
   \label{eq:s2}
\end{equation}
The second term on the left hand side of (\ref{eq:s2}) expands as
\begin{equation}
   < \vfa, \Lr2 \vfa >\; = 
   - \lc < \va{}^2, W'''_0 \vv2 > 
   - \f12 \lc <\va{}^2, W^{IV}_0 (\vv1)^2 >  \, .
   \label{faL2fa}
\end{equation}
The second term is $O(\beta^2 \Lambda^{3})$ while the first term is
$O(\beta^2 \Lambda^{4})$ and dominates at leading order.  This term
can be transformed in the same way as above for $<\vfa,\Lr1 \vfa>$,
leading to
\[
   < \vfa, \Lr2 \vfa > = - \frac{\lc}{\Lambda} \int \Q1 \pxx M_{j} 
   dx + O(\beta^2 \Lambda^{3}) \, ,
\]
where $\Q1$ is the right hand side for the second order version of
(\ref{baseper}).
After dropping out all contributions that vanish owing to 
the symmetries, we obtain
\begin{multline}
   < \vfa, \Lr2 \vfa > = \frac{\beta^2}{\Lambda} \int \pcm \vub \pxx 
   M_{j} dx  \\
   +\frac{\lc}{2 \Lambda} \int W'''_{0} (\vv1)^2 \pxx M_{j } dx 
   + O(\beta^2 \Lambda^{3}) \, .
   \label{faL2fa2}
\end{multline}
The second term on the right hand side of (\ref{faL2fa2}) is negligible  
relatively to the first. This latter can be calculated using 
\[
   \frac{1}{\Lambda} \int \pcm \vub \pxx M_{j} dx = 
   \frac{1}{2 \Lambda} \int_{x_{j}-\f12 \Lambda}^{x_{j}+\f12 \Lambda} 
   \pcm \vub \pxx \vv0 dx = 
   \frac{1}{2 \Lambda} \int_{x_{j}-\f12 \Lambda}^{x_{j}+\f12 \Lambda}
   \pxm \vub \vv0  \, 
\]
and (\ref{vub}). We obtain
\begin{equation} 
   < \vfa, \Lr2 \vfa > = \frac{\beta^2 \Lambda^4 \Gamma ^2}{92,\!160\; s^2 
   \lc}  + O(\beta^2 \Lambda^3) \, .
   \label{faL2fa3}
\end{equation}

The other contributions in (\ref{eq:s2}) involve $\vpa1$ and $\vpb1$. 
These quantities can be approximated in the same way as $\vv1$ at 
distance from the kinks. We have
\begin{equation}
   4 s^2 \lc \vpa1 = - \lc \pxm (W'''_{0} \vv1 \va{}) - \c1 \pxm \vfa 
   + \beta \pcm \vfa - \mu_{1} \pxxm \vfb \, ,
   \label{vpa1}
\end{equation}
and a similar expression for $\vpb1$. Then,
after some algebra and dropping the terms which do not 
contribute to the leading order, we have
\begin{multline*}
   4 s^{2 } \lc <\vfa, \Lr1 \vpa1> = \c1^2 <\va{}, \pxxm \va{}> 
   + \beta^2 <\va{}, \pfm \va{}> \\ - 2 \beta \c1 <\va{}, \pdm \va{}>
   - \mu_{1} \beta <\va{}, \pem \vb{}> + \mu_{1} \c1 <\va{}, \pcm \vb{}>
   + O(\beta^2 \Lambda^{3}) \, .
\end{multline*}
Using the results of Appendix~\ref{s:Green}, we obtain
\[
   <\vfa, \Lr1 \vpa1> = - \frac{\beta^2 \Gamma^2 \Lambda^4}{ 92,\!160\ s^2
   \lc} (1 + 6 t^2 + 15 t^4) + O( \beta^2 \Lambda^{3})\, .
\]
and
\[
   < \vfa, \vpb1 > = - \frac{\beta \Gamma^2 \Lambda^3}{4,\!608\;  s^2 \lc} 
   (t + 3 t^3) + O(\beta \Lambda^2) \, .
\]

Finally
\begin{equation}
   \sigma_2 = - \frac{\beta^2 \Lambda^4}{69,\!120\; s^2 \lc}
   \frac{t^2(4+9t^2)}{(1+t^2)^2} + O(\beta^2 \Lambda^{3}) \, .
   \label{s2r}
\end{equation}
The contribution from $r^{2}$ is a correction to the first order 
stabilization obtained in $\sigma_{1}$. The $\beta$ term does not 
appear at first order and is stabilizing in (\ref{s2r}). Though the 
effect is small, it increases algebraically with $\Lambda$ while the 
non linear coupling of kinks decreases exponentially in  (\ref{s0}).
Therefore, if $r=0$, stabilization of the $m$-mode perturbation to the 
stationary solution is obtained at leading order for 
\begin{equation}
   \beta > \beta_{c} = \left( 35,\!389,\!440\, 
   \frac{e^{-s \Lambda}}{\Lambda^5} s^5 
   \lc^2 \frac{1 }{4 + 9 t^2} \right)^{1/2} \, .
   \label{betac}
\end{equation}
The condition is the most restrictive for $m=1$, that is $t= \tan 
\pi/(2N)$.

\subsection{Stability of $\beta$-CH equation with finite radius of 
deformation}

In this case $\Lr1$ and $\Lr2$ are modified as 
\begin{align}
   \Lr1 &= \lc \px W'''_{0} \vv1 \px + \c1 \px - 
   (\beta + \c1 S^{2})\pxm - r
   \, , \label{L1S:def}  \\
   \Lr2 &=  \lc \px (W'''_{0} \vv2 + \f12 U^{IV}_{0} \vv1^2 ) \px + 
   \c2 (\px - S^2 \pxm) \, .
   \label{L2S:def}
\end{align}

At first order in $\varepsilon$, the imaginary part of the eigenvalue is
\begin{equation}\begin{split}
   \mu_{1 } &= -\c1\frac{<\va{}, \pxm \vb{}>}{<vfb, \pxxm \vfb} +
   (\beta + S^2 \c1) \frac{<\va{}, \pcm \vb{}>}{<vfb, \pxxm \vfb}
   \\
   &= - \frac{4 \beta t (2 + 3 t^2)}{\Lambda S^2 (1 + t^2)} + 
   O\pr{\frac{\beta}{\Lambda^2}}
\end{split}\end{equation}

At second order in $\varepsilon$, $\sigma_{2}$ is still given by 
(\ref{eq:s2}). The contribution $<\vfa, \Lr2 \vfa>$ is now 
\[
   <\vfa, \Lr2 \vfa> = \frac{\beta (\beta + \c1 S^2)}{\Lambda} \int \pcm 
   \vub \pxx M_{j} dx + O\pr{\frac{\beta^2}{\Lambda}} \, .
\]
After integration by part and using (\ref{vubS}), we obtain 
\[ 
   <\vfa, \Lr2 \vfa> = - \frac{\beta^2 \Gamma^2}{40\; s^2 S^4 \lc} 
   + \pr{\frac{\beta^2}{\Lambda}} \, .
\]
Similarly, we have
\begin{align*}
   <\vfa, \Lr1 \vpa1 > &= - \frac{\beta^2 \Gamma^2}{40\; s^2 S^4 \lc}
   (1+6 t^2 + 15 t^4) 
   + O\pr{\frac{\beta^{2}}{\Lambda}} \, , \\
   <\vfa,\vpb1> &= - \frac{\Gamma^2 \Lambda}{96\; s^6 S^{2} \lc} t(1+3t^2)
   +  O(1) \, .
\end{align*}

Finally, we obtain $\sigma_{2}$ as
\begin{equation} 
   \sigma_{2} = - \frac{\beta^2}{30\; s^2 S^4 \lc}
   \frac{3 + 7 t^2 + 9 t^4}{(1 + t^2)^2} 
   + O\pr{\frac{\beta^2}{\Lambda}} \, .
   \label{s2S}
\end{equation}

The stability crossover for $\beta$ is now
\begin{equation}
   \beta_{c} = \left( 15,\!360\, s^{5} S^4 \lc^{2} \frac{3 + 7 t^2 + 9 
   t^4}{t^2}\frac{e^{-s \Lambda}}{\Lambda} \right)^{1/2} \, .
\end{equation}
Therefore, the stabilizing effect of $\beta$ is much reduced compared 
to that with infinite radius of deformation.

All the perturbative calculations of Sections \ref{s:persta}
and \ref{s:stab} have been checked with Mathematica.

\section{Numerical approach}
\label{s:numerical}

The analytic results established in Section~\ref{s:stab} are valid in
the double limit of small $\epsilon$ and large $\Lambda$. These
asymptotic results are complemented and compared with three types of
numerical calculations: (i) numerical solution of the perturbative
problem for several values of $\Lambda$, (ii) direct numerical
simulation of the Cahn--Hilliard equation in the Fourier space and
(iii) direct stability calculation.

\subsection{Numerical solution of the perturbative problem}
\label{s:numpert}

We relax here the hypothesis on the large value of $\Lambda$ by
solving numerically the perturbative equations for $\vv1$, $\vv2$,
$\vpa1$ and $\vpb1$, and numerically evaluating the solvability
conditions.

The calculation is performed according to the following algorithm
\begin{enumerate}
\item The basic solution $\vv0$ is defined by the approximate form
  given in appendix~\ref{aCH}.
\item The inverse derivatives $\pxm \vv0$, $\pxxm \vv0$ and $\pcm
  \vv0$ are calculated by Fourier transform and tabulated to obtain
  $\Q0$.
\item $\c1$ is calculated by discrete evaluation of (\ref{speed}).
\item (\ref{baseper}) is discretized as a tridiagonal problem and
  solved for $\vub$ and $\vur$.
\item $\Q1$ is built in the same way as $\Q0$ from the inverse
  derivatives of $\vub$ and $\vur$.
\item $\vv2$ is calculated by inverting a tridiagonal discretized
  problem.
\item The eigenvectors $\va{}$ and $\vb{}$ are defined using (\ref{defva})
  and (\ref{defvb}).
\item $\vfa$ and $\vfb$ are calculated by Fourier transform and
  tabulated.
\item $\sigma_1$ and $\mu_1$ are calculated according to
  (\ref{eq:s1}), (\ref{eq:m1}) and (\ref{L1:def}).
\item $\vpa1$ and $\vpb1$ are obtained using (\ref{eq:L1}) and solving
  a tridiagonal discretized problem.
\item $\sigma_2$ is obtained from (\ref{eq:s2}), (\ref{L1:def}) and
  ((\ref{L2:def}).
\end{enumerate}

This algorithm admits $O(e^{-s \Lambda/2})$ errors, but all the steps
generating errors of algebraic order in the asymptotic expansion are
here solved numerically. The implementation has been done as a
Mathematica notebook available from the authors. The number of grid
points and Fourier modes has been adjusted as a function of $L$ and
$N$ in order to provide at least 3 digits of accuracy in the results.
The symmetries have been exploited to distinguish the solutions and
reduce the number of points.

\subsection{Numerical solution of the complete Cahn--Hilliard equation}

We have done numerical simulations of the complete Cahn--Hilliard
equation (1) for the case $\gamma=S=0$. For practical convenience, the
spatial period has been kept fixed to $2 \pi$ by rescaling $x$ as $x
\rightarrow p x$. The complete Cahn-Hilliard equation then reads
\begin{equation}
   \partial_T v=
   \frac{\lambda_1}{3 p^2}\partial^2_x v^3-\frac{\lambda_2}{p^2}
   \partial^2_x v - \frac{\lambda_3}{p^4} \partial^4_x v - p \beta 
   \partial_x^{-1} v - r v  \, .
   \label{rbetach}
\end{equation}
Since Fourier modes are discretized by the periodicity condition,
the number $n$ of unstable modes for $v=0$ is the integer part of
$p(2/3)^{1/2}$.

\subsubsection{Time-dependent simulations}
\label{s:tdep}

The simulations are performed using a standard semi-spectral method
where the number of retained real Fourier modes is 256. The
collocation grid in the spatial domain has 512 points in order to
fully remove the aliasing due to the cubic term in (\ref{rbetach}). We
have checked that using higher resolution does not modify the results
within the explored parameter range. The temporal integration is
performed with an Adams-Bashforth second-order scheme.  Initial
conditions are a random white noise in the spatial domain. We have
made a large number of runs by varying the initial conditions
(changing the seed of the pseudo-random number generator and the
amplitude of the noise), the values of $r$ and $n$.

Section \ref{s:stab} shows the existence of multiple stable solutions
induced by $\beta$ and friction but does not provide indications about
the attraction basins of these solutions. In the inverse cascade of
the standard Cahn-Hilliard equation, the interaction of a pair of
neighbor kink and antikink scales as $\exp(-s \Delta x)$, where
$\Delta x$ is the distance between the kink and the antikink (see
Appendix B).  When friction $r$ is just above the critical value
$r_c(N)$ stabilizing the solution with $N$ pairs, we may conjecture
that the stabilizing effect is of the order $O((r-r_c(N))\delta x)$,
where $\delta_x$ is the departure of a kink from its equilibrium
position (this is clear from the shape of the eigenmodes $v_a$ and
$v_b$ (cf. ection \ref{s:CHstab})). This effect, however, cannot
extend very far in $\delta x$ as the attraction to the neighbor
antikink grows as $\exp(s \delta x)$.  Moreover, the time-dependent
solutions do not need to pass in the vicinity of the $N$-pair fixed
point during the cascade of kink-antikink annihilations.  Therefore,
we expect that the fraction of the solutions halting on the $N$-pair
stable state will be small in the vicinity of the critical $r_c$ and
will be significant only when the stabilizing effect is felt over a
distance of order $\Lambda$. Once this is obtained, very few solutions
should jump to lower $N$ states.

\begin{figure}
  \iffigs 
     \centering \includegraphics*[width=10cm]{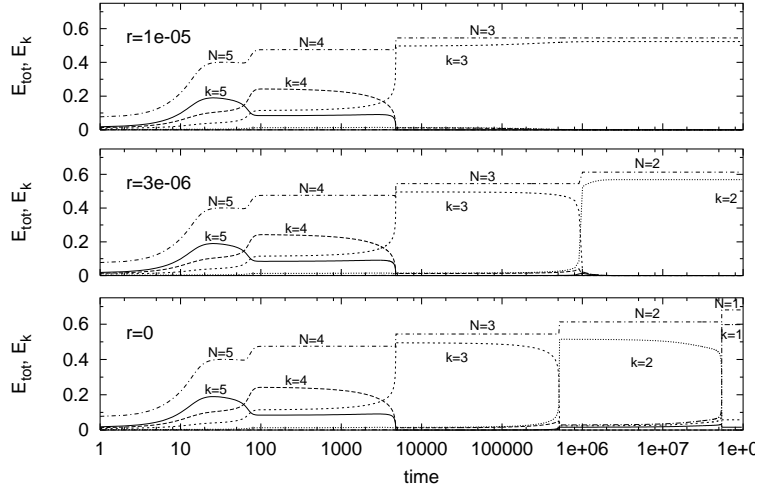}
  \else 
     \drawing 100 10 {Friction simulations} 
  \fi
  \caption {Temporal evolution of the energies $E(k)$ of the Fourier modes
  and of their sum $E_\mathrm{tot}$.  The three shown cases, are for $r=0$,
  $r=3\,10^6$ and $r=10^5$, with $\beta=0$, $n=10$ and the same
  realization of white noise as initial condition.  For $r=0$ the
  inverse cascade is complete to $N=1$; for the other values the
  cascade stops respectively on $N=2$ and on $N=3$. Note that $E_tot$
  is constant between two annihilation events. Increasing further $r$
  enables to stop the cascade on larger $N$ configurations (not
  shown).
  \label{f:FRISIM}}
\end{figure}

\begin{figure}
   \iffigs \centering \includegraphics*[width=10cm]{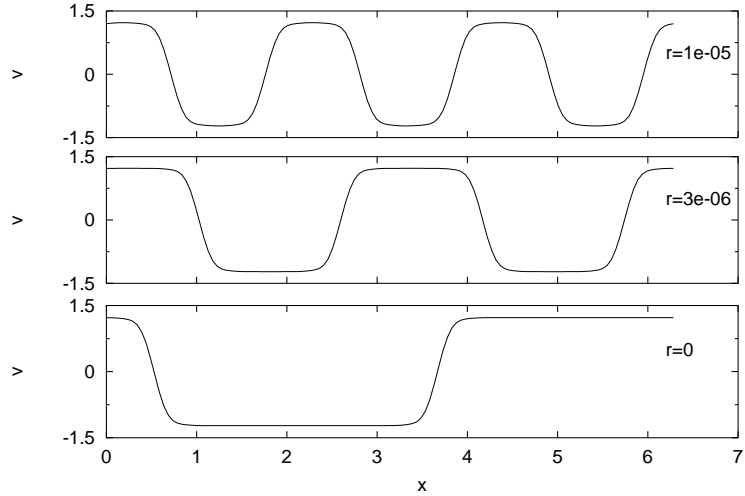}
   \else \drawing 100 10 {Friction ph simulations}
   \fi
   \caption {The corresponding asymptotic velocity profile 
     for the three cases presented in Fig.~\ref{f:FRISIM}.  Note that the
     kinks-antinkinks pairs are equidistant, unlike the $N=2$ or $N=3$
     configurations of Fig.~\ref{f:ANKINK} which are only metastable.
   \label{f:FRISIM_PH}}
\end{figure}

Figure~\ref{f:FRISIM} shows how the inverse cascade evolves as a
function of $r$ for the same initial conditions, with the
corresponding steady states shown in figure~\ref{f:FRISIM_PH}. The
final state wavenumber increases with $r$ and stays bounded by the
value of the most unstable eigenmode of the Kolmogorov flow $k=
(2/3)^{1/2}n$ which is the unique mode excited when $r$ approaches $r_0$.
The total energy calculated as the sum of the various energies E(k) of
the single modes $k$, decreases as the friction increases.

\begin{figure}
  \iffigs \centering \includegraphics*[width=10cm]{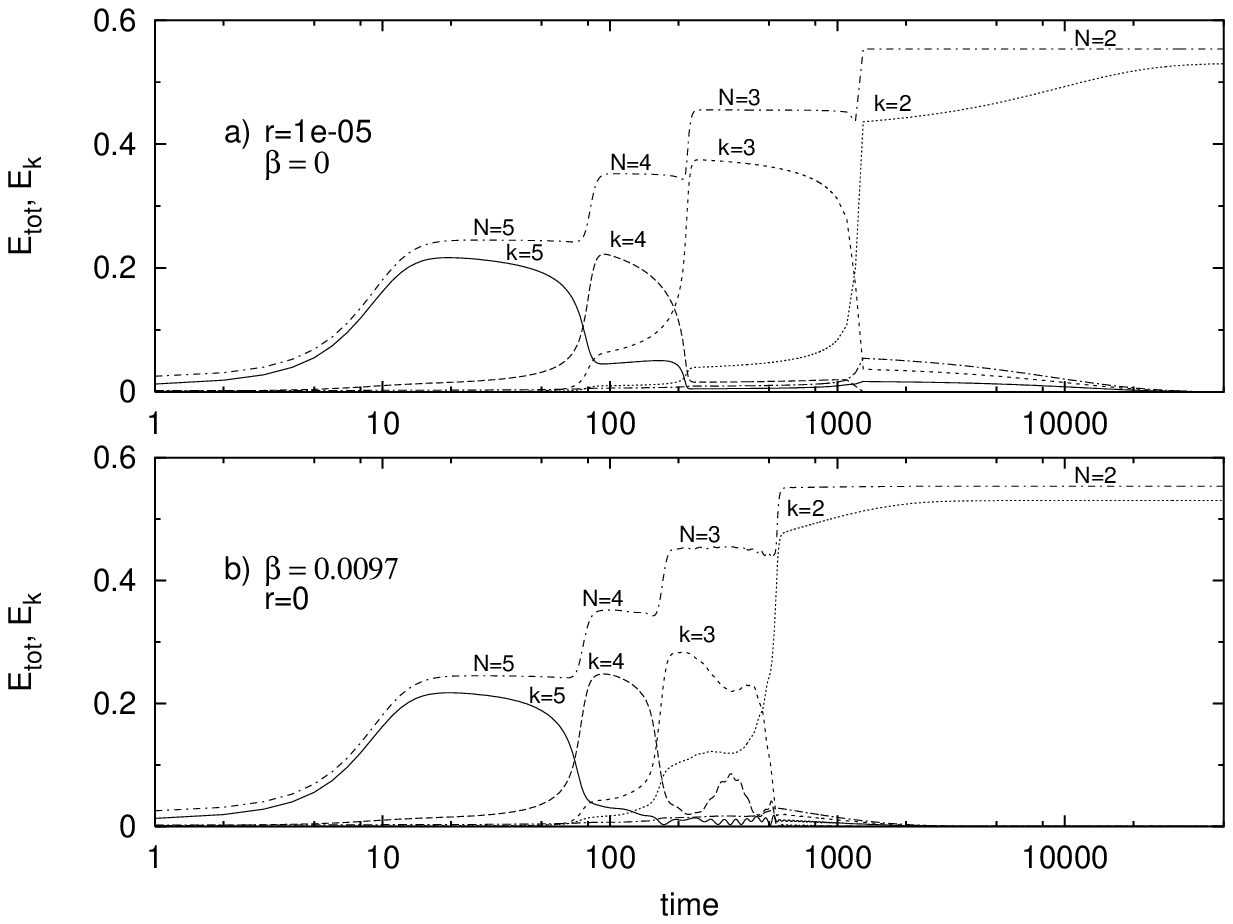}
  \else \drawing 100 10 {Comparison $\beta$ and friction simulations}
  \fi
  \caption {Comparison between two temporal evolution of 
     $E_\mathrm{tot}$ and $E(k)$, for the same initial conditions 
     and for very similar final solutions with same $N$ and final
     energy $E_\mathrm{tot}\approx0.5535$.
     Top: $r=10^{-5}$ and $\beta=0$. Bottom: $r=0$ and $\beta=9.7\,10^{-3}$.
   \label{f:DUOFB}}
\end{figure}

Figure~\ref{f:DUOFB} compares the halting of the inverse cascade by
friction and by $\beta$ effect for the same initial conditions.  It is
apparent from figure~\ref{f:FRISIM} that friction simply halts the
cascade by stabilizing one of the intermediate steps.  The effect of
the $\beta$-term is more complex: oscillatory transients are excited
and, paradoxically, the cascade is accelerated. In Fig.~\ref{f:DUOFB}
the transitions to $N=3$ and to $N=2$ occur much earlier than in the
absence of $beta$. Other examples can be found in
\cite{Frisch:95,Legras:99}.  In the final state, the $\beta$ effect
breaks the symmetry between the kinks and the antikinks which is
preserved by friction.


\subsubsection{Numerical stability calculations}
\label{s:numstab}

In order to compare the results from the analytic and numerical
perturbation approaches to the direct simulation of the complete
Cahn--Hilliard equation, we have developed a direct analysis of
stability by the same technique as for the time-dependent
simulations.  The core of this analysis is to calculate the Jacobian
matrix of the right-hand side of (\ref{rbetach}) linearized around a
given state $v$, with respect to each of the Fourier component.  This
is done by differentiating (\ref{rbetach}) and calculating the columns
of the Jacobian matrix by the semi-spectral method applied to the
differentiated equations for each Fourier component.

The stability calculation is performed as follows.  First an estimate
of the stationary solution based on Appendix~\ref{aCH} is refined by a
Newton-Raphson algorithm which usually converges within a few steps.
The degeneracy due to the $x$-invariance of the complete Cahn-Hilliard
equation is removed by setting to zero the imaginary part of the
dominating Fourier mode and removing the corresponding row and line
from the Jacobian matrix. In the case $\beta \ne 0$, the solution is
stationary in a frame traveling at a velocity $c$. This phase velocity
is treated as an additional unknown increasing the dimension of the
Jacobian matrix by one. The stability of this numerical solution is
then found by finding the eigenvalues and the eigenmodes of the
Jacobian matrix using a standard QR algorithm from LINPACK. In this
procedure, we find both the slow components associated to kink
dynamics and the highly damped modes associated with fast relaxing
transients. The conditioning of the eigenvalue problem gets very bad
as $\Lambda$ increases as a result of the large separation between
slow and fast eigenvalues, thus limiting the parameter range for the
numerical calculation of stability. There is enough overlap, though,
with the validity domain of the perturbative theory to
provide detailed comparison. The number of Fourier modes used in this
analysis has been 128, 256 or 384, depending on the values of $L$ and
$N$.

\subsection{Comparison of stability results}

Table~\ref{t:fricbeta} compares the critical values of friction and
$\beta$ estimated from the analytic perturbative expansion, the
numerical solution to the perturbation problem and the numerical
stability analysis. In the numerical stability analysis the value of
the parameter is adjusted by dichotomy from two values bracketing the
transition.  There is an excellent agreement between the three values
of critical friction when $\Lambda$ is large. At the largest values,
however, the QR algorithm fails to converge and no results are
obtained for the numerical stability.  When $\Lambda$ is not large,
that is when the kinks are not distant enough to neglect the
contribution of $O(e^{-s \Lambda/2})$, significant discrepancies occur
between the different estimates. We see from the table that this
occurs when $e^{-s \Lambda/2} \gtrapprox 3\,10^{-2}$.

There is also an excellent agreement between $\beta_c^\mathrm{pert}$
and $\beta_c^\mathrm{num}$ for the same range of $\Lambda$ values as
for $r_c$.  The analytical prediction (\ref{betac}), however, provides
only an order of magnitude and is wrong by at least a factor two when
the two other quantities agree by four digits. The reason is that the
error in (\ref{betac}) depends algebraically on $\Lambda$ unlike the
error in (\ref{critr}) where the error exhibits an exponential
dependence. Very large values of $\Lambda$ are required to make
(\ref{betac}): for $n=200$ and $N=5$, we obtain
$\beta_c^\mathrm{pert}=2.24\,10^{-42}$ and
$\beta_c^\mathrm{num}=2.35\,10^{-42}$, while for $N=2$ we obtain
$\beta_c^\mathrm{pert}=1.842\,10^{-101}$ and
$\beta_c^\mathrm{num}=1.808\,10^{-101}$.  This difficulty with $\beta$
is entirely due to the need to solve completely the perturbation at
order 1. The phase speed $c_1$ and the frequency $\mu_1$, which are
obtained as solvability conditions at order one, are known with the
same accuracy as $r_c$, as can be checked in Table~\ref{t:cmu}.

\subsection{Comparison of stability properties and time-dependent solutions}

In order to assess the distribution of final states among
the multiple stable steady states we have performed ensemble
simulations for a number of values of $r$ and $\beta$. 
Each numerical integration of the time-dependent numerical model 
described in Section~\ref{s:tdep}
is characterized by $n$, $r$, $\beta$ and the initial condition.
We choose for this latter a white noise with amplitude $A$ in the 
spatial domain. For each value of the parameters and for two
values of the amplitude, $A=0.1$ and $A=1.$, we performed an
ensemble of 100 independent simulations by varying the seed of the random 
number generator. After some time, all simulations 
converge to a final stationary or uniformly traveling state (if $\beta=0$
or $\beta \neq 0$, respectively). The non convergent cases are due
to fast transients leading to nonlinear numerical instabilities.

Table~\ref{t:r} shows the dependence on $r$ when $\beta=0$ and $n=20$.
The distribution of final states agree qualitatively with the analysis
presented in Section~\ref{s:tdep}. Stationary states associated to
given value of $N$ are only reached for $r$ larger than the critical
value $r_c(N)$.  The proportion of solutions reaching these stationary
states is about 15\% for $A=0.1$ and 10\% for $A=1.$ when
$r/r_c(N)\approx 3.$ This proportion grows rapidly as $r$ increases
further while the number of states reaching solutions with smaller $N$
falls dramatically. In practice, it is difficult to find values of $r$
where more than 3 steady states are obtained. It is also visible that
larger amplitude of the initial conditions favor smaller final $N$ and
more dispersion of the final states.

Table~\ref{t:beta} shows the corresponding results as a function of
$\beta$ when $r=0$. They are qualitatively similar to the precedings
although up to 7 different final states are now observed for
$\beta=1.$

\section{Summary and conclusion}
\label{s:conclu}

We have investigated the stabilization induced by friction and $\beta$
effect in the inverse cascade of the large-scale instability of the
Kolmogorov flow.  This problem has been treated by solving
the unidimensional complete Cahn-Hilliard equation
using both numerical simulations and perturbation techniques

In the standard Cahn-Hilliard equation, the kinks and antikinks are
coupled by interactions decreasing exponentially as function of their
separation. The number of kinks decreases with time and their average
separation increases as a result of the inverse cascade. Therefore the
Cahn-Hilliard coupling also decreases with time.

The basic effect of friction is to damp uniformly the motion of the
kinks and antikinks. Nonlinear effects due to the deformation of the
kinks tend to reduce this damping but cannot invert it.  When the
separation is large enough, the damping overcomes the destabilizing
Cahn-Hilliard coupling halting the inverse cascade before it reaches
the gravest mode. The dependence of the critical friction $r_c$ upon
the kink separation $\Lambda/2$ scales as $r_c \sim (e^{-s \Lambda}
/\Lambda)$. The perturbative approach yields a very accurate
analytical expression of $r_c$ because the leading contribution to
$r_c e^{s \Lambda}$, which is algebraic in $\Lambda$ can be obtained
exactly leaving out only terms which are exponentially small in
$\Lambda$.

Stabilization by $\beta$-effect is more complex. It does not contribute 
to the linearization of Cahn-Hilliard equation around a steady state and
appears only at the second order of the perturbative expansion in $\beta$. 
As the first-order step of the perturbation expansion is only solved 
for the leading contribution in a $1/\Lambda$ expansion, the analytic
expression of the critical $\beta_c$ is fairly inaccurate when compared
to numerical solution of the perturbative problem or to direct stability
calculations. Its scaling $\beta_c \sim (e^{-s \Lambda} /\Lambda^5)^{1/2}$ 
is, however, correctly predicted.

The presence of a finite radius of deformation $1/S$ , which slows
down the fastest Rossby waves, is to provide less efficient
stabilization by the $\beta$ effect.  The critical $\beta_c$ then
scales as $\beta_c \sim (e^{S^4 -s \Lambda} /\Lambda)^{1/2}$.

In the presence of mean advection, as in Ref.~\cite{Manfroi:99}, the
eastward jets are narrow and strong while the westward jets are broad
and slow.  The critical value of friction is reduced (see
(\ref{eq:rcadv})) and scales as $r_c \sim (e^{-s
  \Lambda(1-\gamma/\Gamma)} /\Lambda)$

Stabilization is demonstrated near the modified steady states of the
Cahn-Hilliard equation. The attraction basin of these steady states
depends on the stability of time-dependent solutions and has been
investigated numerically. The results suggest a fairly simple pattern
where, for most values of the parameters, the phase-space is filled by
the attraction basins of only 2 or 3 stationary solutions. The
boundary of these basins might be very complicated, even fractal.

We have observed in the numerical simulations of the pure
Cahn-Hilliard equation that the inverse cascade does not always begin
by the most unstable state $N=k_m$; however it is in principle always
possible to stop the cascade at such a state by enhancing the
friction. The same is not true for $\beta$, which is not always able
to stop the cascade at the scale corresponding to the most unstable
state. As already noticed, $\beta$ dispersive effect is more complex
than friction effect. While the inverse cascade prior to the halting
by friction does not differ from the pure Cahn-Hilliard case, the
$\beta$ effect is paradoxically accelerating the cascade before
halting. We speculate that this is due to the propagation of fast
Rossby waves superimposed to the kinks and increasing their coupling.

Our result for the critical friction $r_c$ differs from that given in
Ref.~\cite{Manfroi:99} where the authors found $r_c \sim
\Lambda^{-3}$.  Their reasoning was based on varying and minimizing
the Lyapunov functional with respect to $\Lambda$. Our interpretation
is that this is questionablesince the Lyapunov functional can only be
minimized with respect to the solution, not with respect to the
parameters.

The scaling of the critical $\beta_c$ also differs from the scaling
$\beta_c \sim \Lambda^{-3}$ which would arise from standard
phenomenology \cite{Rhines:75} by balancing the nonlinear and the
dispersive term in (\ref{vbetach}). The reason lies in the suppression
of nonlinearities in the slow manifold for solutions of the complete
Cahn-Hilliard equation once the initial transients have been
dissipated. In more realistic two-dimensional or quasigeostrophic
flows, the presence of strong dominating jets and/or coherent eddies
is similarly inducing a reduction of nonlinearities with respect to a
plain dimensional estimate. This is why the prediction of a $k^{-3}$
energy spectra \cite{Kraichnan:67} is usually not observed in forced
two-dimensional flows \cite{Legras:88} with the possible exceptions of
the smallest quasi-passive scales of the motion.

\section*{Acknowledgments}

We thank Joanne Deval for her careful checking of the perturbation 
calculations. We thank Uriel Frisch for his encouragements and numerous
discussions.

\newpage
\appendix

\setcounter{section}{0}
\section{Approximate solution of the Cahn-Hilliard equation}
\label{aCH}
\setcounter{equation}{0}

The stationary form of the Cahn--Hilliard equation $\pxx v - U'(v)=0$
can be recasted as
\begin{equation}
   \px U = \f12 \px (\px v)^{2} \,,
   \label{statiobase}
\end{equation}
from which the solution is obtained by quadrature.
By integrating (\ref{statiobase}) once, we obtain
\begin{equation}
U = \f12 (\px v)^{2} -C \,,
\label{statioquad}
\end{equation}
where $C$ is a constant which determines the value of $\px v$ on the 
kinks and the periodicity of the solution. For the single-kink 
solution (\ref{kink}), we have $C= \f12 s^{2} \Gamma^{2}$ and the 
asymptotic value of $U$ is $-C$. 
Let us now assume 
\[ C = \f12 s^2 \Gamma^2 (1 - \mu) \, , \]  
where $\mu$ is assumed a small perturbation and try a 
solution under the form
\begin{equation} 
   v = \Gamma \tanh s x + \mu \tv \, .
   \label{approxv}
\end{equation}
By replacing in (\ref{statioquad}) and expanding, we obtain, at first 
order in $\mu$,
\begin{equation}
   2 \px \tv = -4 s \tv \tanh s x - \Gamma s \cosh^2 sx \, .
   \label{vtequ}
\end{equation}
This equation can be solved as
\begin{equation}
   \tv = - \frac{\Gamma}{16} \pr{(3 + \cosh^{2} s x)\tanh s x + \frac{3 s 
   x}{\cosh^2 s x}} \, ,
   \label{vtsol}
\end{equation}
where the condition $\tv(0) = 0$ has been used. We can relate 
$\mu$ to the period of the solution by using $\px v = 0$ for $x = 
\Lambda/4$, leading to
\begin{multline}
   \mu\left\{\f18 s \Gamma \pr{\tanh^{2} sx(3+2\cosh^{2}sx) + 
   \frac{3s \tanh s x}{\cosh^{2} sx}} \right. \\ \left.
   - \f12 \Gamma s \cosh^{2} s x \right\} + s \frac{s 
   \Gamma}{\cosh^2 s x} = 0 \, .
   \label{musol0} 
\end{multline}
When $\Lambda$ is large the main contribution is 
\begin{equation}
   \mu = 64 e^{-s \Lambda} \,.
   \label{musol}
\end{equation}

Comparison with numerical solutions of (\ref{vbetach}) shows that
(\ref{approxv}) with (\ref{vtsol}) and (\ref{musol}) approximates
periodic stationary stationary to less than 0.2\% for $\Lambda>10$.
One can build a solution over the whole domain by using
(\ref{approxv}) over contiguous intervals containing a kink matched at
mid-distance between adjacent kinks.  The approximate solution is
continuous and the discontinuity of its derivative at matching points
is $O(\exp(-s |x_{AK}-x_K|))$ where $|x_{AK}-x_K|$ is the distance
between adjacent kinks.

\setcounter{section}{1}
\section{Kink motion in the Cahn-Hilliard equation}
\label{s:CH}
\setcounter{equation}{0}

This Section is adapted from Kawasaki \& Ohta~\cite{Kawasaki:82} and
corrects one error found in this paper.

We assume that the solution is a combination of kinks which are
individually described by (\ref{kink}). As $\pxx M_{\ell}(x)$ and
$W'(M_{\ell})$ decay rapidly away from $x=x_{\ell}$, the solution in
the vicinity of the $j$-th kink is the sum of $M_j(x)$ and of small
contributions from adjacent kinks which are the small deviations from
their asymptotic values. We write
\begin{equation}
   v(x,t) = M_j(x) + \tilde{v}_j(x,t) \, ,         
\end{equation}
where $\tilde{v}_j$ is small in the vicinity of the $j$-th kink, but
takes finite value at distance. A valid expression in the vicinity of
neighbor kinks is
\begin{equation}
   \tilde{v}_j(x,t) = \sum_{\ell<j} (M_{\ell}(x) - M_{\ell}(+ \infty))
   + \sum_{\ell>j} (M_{\ell}(x) - M_{\ell}(- \infty)) \, ,
\end{equation}
where $M_{\ell}(+ \infty)$ and $M_{\ell}(- \infty)$ are the 
asymptotic values at infinity for the basic kink profile.
Here we assume that kinks and antikinks alternate(i.e. $\epsilon_j 
\epsilon_{j+1} = -1$), and that they are numbered from $0$ to $2N-1$ 
within the periodic interval $[0,L]$. The time dependence is entirely 
contained within the positions of the kinks $\{x_j(t)\}$.

The temporal evolution of the solution is then given by:
\begin{equation}
   \pt v(x,t) = - \sum_{\ell=0}^{2N-1} \dot{x}_{\ell}(t) \px M_{\ell}(x) \,,
   \label{vtemp}
\end{equation}
where $v(x,t)$ is governed by 
\begin{equation}
   - \frac{1}{\lc} \pxxm \pt v = 
   \pxx v - W'(v) + h(t) 
   \label{CH}\, .
\end{equation}

The function $h(t)$ arises from the integration in $x$. The other 
terms arising from the integration vanish owing to the periodicity 
in $x$.

In order to estimate the motion of the $j$-th kink, 
we use $M'_j$ as a test function by multiplying (\ref{CH}) 
and integrating over the domain. Then we obtain:
\begin{equation}
   \begin{array}{ccccc}
      \underbrace{\int_0^L \frac{1}{\lc} \sum_{\ell=0}^{2N-1} 
      \dot{x}_{\ell} \px M_j \pxm M_{\ell} dx} &
      = & \underbrace{\int_0^L ( \pxx v - W'(v)) \px M_j dx} & + 
      &\underbrace{\int_0^L  h \px M_j dx} \, . \\
      A&&B&&C 
   \end{array}
   \label{kmoto}
\end{equation}

Contribution $A$ can be written as
\begin{equation} 
   A = -\frac{1}{\lc} 
   \sum_{\ell=0}^{2N-1} \dot{x}_{\ell} \int_0^L \int_0^L \px M_j(x) 
   \Gr2(x-x') \px M_{\ell}(x') dx dx'\,,
   \label{A0} 
\end{equation}
where $\Gr2$ is the Green function solution of 
\[ - \pxx \Gr2(x) = \delta(x) \, .\]
$\px M_j$ and $\px M_{\ell}$ are two well separated functions which 
contribute to the integral in (\ref{A0}) respectively in the close vicinity of 
$x_j$ and $x_{\ell}$. 
By expanding $\Gr2(x-x')$ near $\Gr2(x_j-x_{\ell})$ and summing local 
contributions using 
$\int (x-x_j)^2 \px M_j dx = (-1)^j \pi^2 \Gamma / 6 s^2$ and
$\int\int |x-x'| \px M_j \px M_{\ell} dx dx' = 4 \Gamma^2 /s$, we obtain
\begin{equation}
   A = -\frac{4 \Gamma^2}{\lc} 
   \sum_{\ell=0}^{2N-1} \dot{x}_{\ell}
   \pr{(-1)^{j-\ell}\Gr2(x_j-x_{\ell})+ (-1)^{j-\ell} \frac{\pi^2}{12 L s^2}
   - \frac{1}{2 s}\delta_{j-\ell}} \, .
\end{equation}  
Using the expression for $\Gr2$ within the interval $[0,L]$, one
obtains
\begin{multline}
   A = -\frac{2 L \Gamma^2}{\lc} \sum_{\ell=0}^{2N-1} (-1)^{j-\ell}
   \dot{x}_{\ell}
   \left( \frac{((x_j-x_{\ell})[L])^2}{L^2} - 
   \frac{(x_j-x_{\ell})[L]}{L} \right. \\ \left. 
   + \f16 + \frac{\pi^2}{24 L^{2} s^2}
   - \frac{1}{4 L s} (-1)^{j-\ell} \delta_{j-\ell} \right) \,.
   \label{A2}
\end{multline}

Contribution $B$ is expanded using 
\begin{equation}
   W'(v) = W'(M_{j}) + W''(M_{j}) \tilde{v}_{j} + 
   W'_{NL}(M_{j},\tilde{v}_{j}) \, .
\end{equation}
Like $\tilde{v}_{j}$, $W'_{NL}$ is small in the vicinity of the $j$-th 
kink but finite at distance. We have
\begin{multline}
   B = - \int_{O}^{L} W'_{NL} \px M_{j} dx
   + \int_{O}^{L}  (\pxx M_{j} - W'(M_{j})) \px M_{j} dx \\
   + \int_{O}^{L} (\pxx \tilde{v}_{j} - W''(M_{j}) \tilde{v}_{j}) 
   \px M_{j} dx \, .
   \label{eq:B}
\end{multline}
The second integral in (\ref{eq:B}) vanishes and the third one 
vanishes also after integration of its first term by part. Therefore, 
we are left with
\begin{equation} 
   B = - \int_{O}^{L}  W'_{NL} \px M_{j} dx \, .
   \label{eq:B2}
\end{equation}
Using (\ref{Wdef}) we find
\[ W'_{NL} = \frac{2 s^2}{\Gamma^2} \tilde{v}_{j}^2 (3 M_{j}  + 
\tilde{v}_{j}) \, . \]
In the vicinity of $x_{j}$, $\tilde{v}_{j}$ is of the order of the 
tails of $M_{j+1}(x_{j})-M_{j+1}(- \infty)$ and 
$M_{j-1}(x_{j})-M_{j-1}(\infty)$, that is $O(\max(e^{-2s (x_{j+1}-x_{j})},
e^{-2s (x_{j}-x_{j-1})})$.
In the vicinity of $x_{j+1}$ and $x_{j-1}$, $\tilde{v}_{j}$ is $O(1)$ 
while $\px M_{j}$ is $O(e^{-2s (x_{j+1}-x_{j})})$ and $O(e^{-2s (x_{j}-x_{j-1})})$.
Therefore, the two main 
contributions to B in (\ref{eq:B2}) arise from the vicinities of 
$x_{j+1}$ and $x_{j-1}$. In the vicinity of $x_{j+1}$ we use
\[ \tilde{v}_{j} = M_{j+1} + \epsilon_{j+1} \Gamma \]
so that 
\[ \tilde{v}_{j}^2 (3 M_{j}  + \tilde{v}_{j}) = \epsilon_{j} \Gamma^3
(2 - \tanh s (x-x_{j+1}))(1+\tanh s (x-x_{j+1}))^2 \, .
\]
Using also
\[ \px M_{j} = 4 s \epsilon_{j} \Gamma e^{-2s(x-x_{j})} \, , \]
and replacing in (\ref{eq:B2}) with similar contributions from the 
vicinity of $x_{j-1}$, we obtain 
\begin{equation}
   B = - 32 s^2 \Gamma^2 (e^{-2s(x_{j+1}-x_{j})} 
   -e^{-2s(x_{j}-x_{j-1})}) \, .
\end{equation}
        
Finally, contribution $C$ gives
\begin{equation}
   C = 2 \epsilon_j \Gamma h(t) \, .
\end{equation} 

Summarizing the results, one gets
\begin{multline}
   \label{motok}
   \frac{2 L \Gamma^2}{\lc} \sum_{\ell=0}^{2N-1} (-1)^{j-\ell}
   \left( \frac{((x_j-x_{\ell})[L])^2}{L^2} - 
   \frac{(x_j-x_{\ell})[L]}{L} \right. \\ \left. 
   + \f16 + \frac{\pi^2}{24 L^2 s^2}
   - \frac{1}{4 L s} (-1)^{j-\ell} \delta_{j-\ell} \right) 
   \dot{x}_{\ell} \\ =
   32 s^{2} \Gamma^2 \pr{e^{-2s(x_{j+1}-x_j)} - 
   e^{-2s(x_j-x_{j-1})}} - 2 \epsilon_j \Gamma h(t) \,.
\end{multline}

Eq. (\ref{motok}) shows that two neighbor kink and antikink attracts
themselves A stationary solution is obtained when $B$ vanishes for all
values of $j$. This condition is satisfied if the kinks and antikinks
are equispaced over the interval $[0,L]$. Then, $h(t)=0$.

Eq. (\ref{motok}) can be used to determine the motion of the kinks
far from the equilibrium, under the condition that the kinks and 
antikinks remain far enough to satisfy the approximations of the above 
calculation.

The calculation of Kawasaki \& Ohta ~\cite{Kawasaki:82} slightly differs from 
our own and is limited to the leading order. They 
fail to take into account the exponential variation of 
$\tilde{v}_{j}$ near $x_{j+1}$ and $x_{j-1}$. Therefore, their result 
for the leading order of $B$ contains an error, being too small by a factor 2. 

\setcounter{section}{2}
\section{Green function in the periodic domain and calculations of 
coupling coefficient}
\label{s:Green}
\setcounter{equation}{0}
Within the periodic domain $[0,L]$, the $\delta$ function is made periodic
by adding a constant value $-1/L$ everywhere but at the origin.
One can also use
\[ 
   \delta(x) = \f1L \sum_{n \neq 0} \exp{\left(i \frac{2 \pi n}{L}x\right)} 
\, . \]
The solution to 
\[
   \partial^n_{x} \mathcal{G}_n(x) = - \delta(x)
\]
is 
\[    
   \Gr{n}(x)  =  L^{n-1} g_n \pr{\f{x}{L}}  \, ,
\]
where
\begin{align*}
    g_1(x) &= x[1] - \f12 \,, \\
    g_2(x) &=  \f12 ((x[1])^2 - x[1] + \f16)  \,, \\
    g_3(x) &= \f14\pr{\f23 (x[1])^3 - (x[1])^2 +\f13 x[1]} \,, \\
    g_4(x) &=  \f1{24}\pr{ (x[1])^4 -2 (x[1])^3 + 
      (x[1])^{2} -\f1{30}} \,, \\
    g_5(x) &= \f1{720} \pr{6( x[1])^5 - 15 (x[1])^4 + 10(x[1])^3
      -x[1]} \,, \\
    g_6(x) &= \f1{720}\pr{ (x[1])^6 -3 (x[1])^5 + \f52 (x[1])^4 
      -\f12  (x[1])^{2} + \f1{42}} \,,
\end{align*}
where $x[1]$ means $x$ modulo 1.

The calculation of the perturbed motion requires to calculate the
Fourier transform
\[        
   \FGr{n}(m) \equiv \sum_{j=0}^{2N-1} (-1)^{j} \Gr{n} 
   \pr{\frac{j \Lambda}{2}} e^{-i \pi \frac{jm}{N}}\, , 
\]
which can be written as a function of
\[        
   s_m(p) \equiv \sum_{j=0}^{2N-1} (-1)^{j} \pr{\frac{j}{2N}}^p
   e^{-i \pi \frac{jm}{N}} \,.
\]
For $m \neq N$, $s_m(p)$ can be calculated using the following 
relations :
\[ 
   P(z,x,J,p) \equiv \sum_{j=0}^{J-1} j^p z^{jx} = \pr{\frac{1}{\ln z}}^p 
   \partial_{x}^p \sum_{j=0}^{J-1} z^{jx} \, , \]
   \[ s_m(p) = \pr{\frac{1}{2N}}^p P(-e^{-i\frac{\pi m}{N}},1,2 N,p) \,. 
\]
We get
\begin{align*}
   s_m(0) &=   0 \,,\\
   s_m(1) &=  -\frac{1}{1+a}  \,,\\
   s_m(2) &=  -\frac{1}{1+a} + \frac{a}{N(1+a)^2}  \,,\\
   s_m(3) &=  -\frac{1}{1+a} + \frac{3a}{2N(1+a)^2}
     + \frac{3a(1-a)}{4N^2(1+a)^3} \,, \\
   s_m(4) &=  -\frac{1}{1+a} + \frac{2a}{N(1+a)^2}
     + \frac{3a(1-a)}{2N^2(1+a)^3} 
     + \frac{a(1-4a+a^2)}{2N^3(1+a)^4} \,, \\
   s_m(5) &=  \cdots + \frac{5a(1-11a+11a^2-a^3)}{16 
     N^{4}(1+a)^{5}} \,, \\
   s_m(6) &=  \cdots + \frac{3a(1-26a+66a^2-26a^3+a^4)}
     {16 N^5(1+a)^6} \,,
\end{align*}
where $a= \exp(-i \theta_m)$ with $\theta_m = \pi m/N$.
We also have
\begin{align*}
   s_N(0) &=  2N \,,\\
   s_N(1) &=  N - \f12  \,,\\
   s_N(2) &=  \frac{1}{12N}(2N-1)(4N-1) \,,\\
   s_N(3) &=  \frac{1}{8N}(2N-1)^2 \,, \\
   s_N(4) &=  \frac{1}{240N^3}(2N-1)(4N-1)(12N^2 -6N -1) \,, \\
   s_N(5) &=  \frac{1}{96N^3}(2N-1)^2(8N^2 -4N -1)  \,, \\
   s_N(6) &=  \frac{1}{1344N^5}(2N-1)(4N-1)(48N^4 -48N^3+6N+1) \,.
\end{align*}
Using these relations, the Fourier transforms are readily calculated.
For $m \neq N$, we have
\begin{align}
   \FGr1(m) 
   &= s_m(1) - \f12 s_m(0) + \f12 \; = \; - \f12 i t
   \label{sf} \,,\\
   \FGr2(m) 
   &=  \frac{L}{2} \pr{ s_m(2) - s_m(1) + \f16 s_m(0)}
   \; = \; \frac{\Lambda}{8} (1 + t^2)
   \label{sg} \,,\\
   \FGr3(m) 
   &=  \frac{L^2}{4} \pr{ \f23 s_m(3) - s_m(2) + \f13 s_m(1)}
   \; = \;  i \frac{\Lambda^2}{32} t (1+t^2)
   \label{sh} \,,
\end{align}
with $t = \tan \pi m / 2 N $.
Since the Fourier transform $\FGr{n}$ scales as $\Lambda^{n-1}$, it 
depends only on the $O(1/N^{n-1})$ term in $s_{m}(n)$. Therefore, the
higher order transforms are, for $m<N$ :
\begin{align}
   \FGr4(m) 
   &=  - \frac{\Lambda^3}{384} (1+t^2) (1+3t^2)
   \label{si} \,,\\
   \FGr5(m) 
   &=  -i \frac{\Lambda^4}{1,\!536} t(1+t^2) (2+3t^2)
   \label{sj} \,,\\
   \FGr6(m) 
   &=  \frac{\Lambda^5}{30,\!720} (1+t^2) (2+15t^2+15t^4)
   \label{sk} \,.
\end{align}

Using this formalism, it is possible to calculate $<\va{}, 
\partial_{x}^{-n} \va{}>$ for even $n$ as
\begin{multline} \label{vanvab}
   < \va{}, \partial_{x}^{-n} \va{}> 
   \, = \, - \frac{1}{L} \int_{0}^L \int_{0}^L \va{}(x) 
   \Gr{n}(x-x') \va{}(x') dx dx'  \\
   = - \frac{1}{L} \sum_{j=0}^{2N-1} \sum_{l=0}^{2N-1}
   \cos j \theta_{m} \cos l \theta_{m}
   \int _{0}^L  \int_{0}^L  \px M_{j}(x) \px M_{l}(x') 
   \Gr{n}(x-x') dx dx' \,.  
\end{multline}
This integral contains two types of contributions. Type I arises from 
the interaction of distant kinks and type II is a local correction 
arising from the autocoupling of a given kink and taking into account 
the discontinuity of $\Gr{n}(x)$ or its derivatives in $x=0$. At first, 
we consider only the type I contribution for which $\Gr{n}(x-x')$ can 
be developed as a Taylor series around $x_{j}-x_{l}$ in order to 
separate the double integration in two independent integrations 
around the kinks
\begin{multline*}
   \Gr{n}(x-x') = \Gr{n}\pr{\frac{(j-l) \Lambda}{2}} + 
   \Gr{n-1}\pr{\frac{(j-l) \Lambda}{2}}
   ((x-x_{j})-(x'-x_{l})) \\ 
   + \f12 \Gr{n-2}\pr{\frac{(j-l) \Lambda}{2}}
   ((x-x_{j})-(x'-x_{l}))^2 + \cdots \, .
\end{multline*}
Now we replace in (\ref{vanvab}) and calculate the local contributions 
using
\begin{align*}
   \int \px M_{j} dx  &= 2 \Gamma (-1)^j  \,,\\
   \int (x-x_{j})\px M_{j}  dx &= 0  \,,\\
   \int (x-x_{j})^2 \px M_{j} dx &= \frac{\pi^2 \Gamma}{6 s^2} (-1)^j \,.
\end{align*}
We also expand the trigonometric factor and find that the only 
non vanishing contribution depends on $j-l$. After relabeling, we 
obtain
\begin{multline*}
   <\va{}, \px^{-n} \va{}>_{I} = - \frac{4 \Gamma^{2}}{\Lambda}
   \sum_{j=0}^{2N-1} \cos \frac{\pi m j}{N} (-1)^j \Gr{n}\pr{\frac{j 
   \Lambda}{2}} \\
   - \frac{\pi^2 \Gamma^{2}}{3 s^2 \Lambda}
   \sum_{j=0}^{2N-1} \cos \frac{\pi m j}{N} (-1)^j \Gr{n-2}\pr{\frac{j 
   \Lambda}{2}} \,,
\end{multline*}
that is
\begin{equation}
   <\va{}, \px^{-n} \va{}>_{I} = - \frac{4 \Gamma^{2}}{\Lambda} 
   \FGr{n}(m) - \frac{\pi^2 \Gamma^{2}}{3 s^2 \Lambda} \FGr{n-2}(m) 
   + O (\Lambda^{n-6})
   \label{vanva}
\end{equation}
for even $n$.
A similar relation is obtained for odd $n$ :
\begin{equation}
   <\va{}, \px^{-n} \vb{}>_{I} = - \frac{4 \Gamma^{2}}{\Lambda} 
   \f1i \FGr{n}(m)  - \frac{\pi^2 \Gamma^{2}}{3 s^2 \Lambda} \f1i \FGr{n-2}(m) 
   + O (\Lambda^{n-6}) \,.
   \label{vanvb}
\end{equation}

The local type-II contributions must be examined case by case. For 
$n=2$, $\Gr2(x)$ is continuous in $x=0$ but its derivative is not. 
Near $x=0$, we have 
\[ 
   \Gr2(x) = \frac{L}{2} \pr{\f16 - \left| \frac{x}{L}\right| + 
   \frac{x^2}{L^2}} 
\]
The two terms $\f16 + \frac{x^2}{L^2}$ are taken into account in type-I 
contribution. The complementary contribution is
\[
   <\va{}, \pxxm \va{}>_{II} =  \frac{1}{2 \Lambda} \int \int 
   \px M(x) \px M(x') | x-x'| dx dx' \,.
\]
After a bit of algebra, we obtain 
\[
   <\va{}, \pxxm \va{}>_{II} =  \frac{2 \Gamma^2}{\Lambda s}
\]
and finally
\begin{equation}
   <\va{}, \pxxm \va{}> = - \frac{\Gamma^2}{2}(1+t^{2}) + 
   \frac{2 \Gamma^2}{\Lambda s} \, .
\end{equation}
For $n=4$, $\Gr4(x)$ has a discontinuity on its third derivative in 
$x=0$. It brings an $O(1/\Lambda)$ correction in $<\va, \px^{-4} 
\va >$, which is of higher order than the terms in (\ref{vanvb}).

For odd orders, including $n=1$, the sine factor cancels the type-II
contribution. Notice that for $n=1$, $\Gr1(x)$ is discontinuous in
$x=0$.  We have arbitrarily assumed that $\Gr1(0)=0$ in (\ref{vanvab})
but this is not important since only the imaginary part of $\FGr1(m)$
is used in the sine transform.

Some other quantities may need to be calculated, of the type $<f
\va{}, \px^{-n} \va{}>$ where $n$ is odd and $f$ is a period-$\Lambda$
non localized function which is odd over the kinks. In this case, the
non vanishing contributions are those arising from the odd derivatives
of $\Gr{n}$. At leading order, we obtain
\begin{equation}
   <f \va{}, \px^{-n} \va{}> = - \frac{2 \Gamma}{\Lambda} \FGr{n-1} (m) \int x 
   f(x) \px M(x) dx + \cdots \, .
   \label{vanfva}
\end{equation}
The same expression holds when $\va{}$ is replaced by $\vb{}$.
The case $n=1$ is special, we have
\[
   <f \va{}, \pxm \va{}> = \frac{1}{2 \Lambda} \int  
   f(x) \px M^2(x) dx \, .
\]

\setcounter{section}{3}
\section{The effect of mean advection}
\label{s:meanadv}
\setcounter{equation}{0}

The case $\gamma \neq 0$ has been studied numerically and near the
linear limit $r_0$ (see Section \ref{s:CHstab}) in
Ref.~\cite{Manfroi:99}. It is possible to study the stability
properties of steady solutions for small $r$ in the same way as for
$\gamma=0$ but this is to the price of a considerable increase of
complexity in the algebra. It is beyond the scope of this manuscript
to describe the details of these cumbersome calculations.  We provide
here the results without demonstration.

When $\gamma \neq 0$, the equilibrium positions of the kinks and antikinks
are given by
\begin{align*}
  x_{2p} &= p \Lambda - \frac{\Delta}{4}  \, , \\
  x_{2p+1} &= (2p+1) \frac{\Lambda}{2} + \frac{\Delta}{4} \, , 
\end{align*}
where $\Delta$ is related to $\gamma$ by
\begin{equation} 
  16 \Gamma e^{-s \Delta} \sinh s \Delta = - \frac{\Delta \Gamma}{\Lambda} 
  + \gamma \, .
\end{equation}
We define also $d=\Delta/\Lambda$.

Similarly to Section~\ref{s:CHstab}, it is convenient to expand the
displacements of the kinks with respect to the equilibrium in terms
of Fourier components. One has now to separate the kinks and
antikinks as 
\begin{align*}
 \delta x_{2p}      &= \sum_{m=0}^{N-1} \psi_m^- e^{i \pi\frac{2mp}{N}} \, ,\\
 \delta x_{2p\!+\!1}&= \sum_{m=0}^{N-1} \psi_m^+ e^{i \pi\frac{(2p\!+\!1)m}{N}}
 \, .
\end{align*}
By combining these components into 
\[
  \Phi_m = \frac{1}{\sqrt{2}} 
  \left( \begin{array}{cc}
    1 & 1 \\
    e^{-i \theta_m} & e^{i \theta_m}
  \end{array} \right)
  \left( \begin{array}{c}
    \psi_m^-\\
    \psi_m^+
  \end{array} \right) \, ,
\]
we obtain
\begin{equation}
  \dot{\Phi_m}=                           
  \frac{\mathcal{A}\sin^2\theta_m}{\mathcal{E}_m} 
  \left(\begin{array}{cc}
    1-\mathcal{C}\cosh s\Delta - \mathcal{Q}_m & 0 \\
    0 & 1-\mathcal{C}\cosh s\Delta +\mathcal{Q}_m
  \end{array} \right) \Phi_m \,
  \label{eq:Phieq}
\end{equation}
with
\begin{align*}
  \mathcal{A}  &=\frac{128\lc \,e^{-s\Lambda}}{(1-d^2)\,\Lambda} D , \\
  \mathcal{E}_m&=1-\frac{4}{s\Lambda(1-d^2)} +\frac{4\sin^2\theta_m}
    {s^2\Lambda^2(1-d^2)} , \\
  \mathcal{B}  &=\frac{d\cosh s\Delta-\sinh s\Delta}{D} , \\
  \mathcal{C}  &=\frac{2}{s\Lambda D} , \\
  D &= \cosh s\Delta - d \sinh s\Delta , \\
  \mathcal{Q}_m &=\left(\mathcal{B}+2\mathcal{C} \sinh s\Delta 
    +\frac{1}{2}\mathcal{C}^2(\cosh 2s\Delta+\cos 2\theta_m)\right)^{1/2} .
\end{align*}
This result generalizes (\ref{s0}).

Now, the stabilization effect by friction is still given at first order 
of the perturbative expansion for small $r$. 
We first need to define
\[
  \rho_m = 2(\mathcal{B} \cos^2 \theta_m + \mathcal{Q}_m \sin^2 \theta_m)^{1/2}
\]
\[\begin{array}{ll}
  \cos \phi_{1m} = \frac{2}{\rho_m} (\mathcal{B} \cos \theta_m) ,
& \sin \phi_{1m} = \frac{2}{\rho_m} (\mathcal{Q}_m \sin \theta_m) , \\
  \cos \phi_{2m} = \frac{2}{\rho_m}(\mathcal{B} + \mathcal{C} \sinh s \Delta
    \sin^2 \theta_m) ,
& \sin \phi_{2m} = -\frac{1}{\rho_m}
    (\mathcal{C} \sin 2 \theta_m \cosh s \Delta).
\end{array}\]
The stabilization effect is 
\begin{multline}
  \sigma_1 = -r + \frac{r \sin^2 \theta_m}{2}
  \left(1-d^2 - \frac{4}{s \Lambda}\right) \\ \times
  \left(1 + d \sin \theta_m \sin (\phi_{2m} - \phi_{1m})
  - \cos \theta_m \cos (\phi_{2m} - \phi_{1m}) 
  - \frac{2 \sin^2 \theta_m}{s \Lambda}\right)^{-1} \, .
  \label{eq:s1adv}
\end{multline}

These results have been checked numerically with the same accuracy as those
presented for $\gamma=0$.

In the case of large $\Lambda$ and when $d$ is not small, that is when
the asymmetry is strong, we can neglect all terms $O(e^{-s \Delta})$ in
front of terms which are $O(1)$ or larger. Equations (\ref{eq:Phieq})
and (\ref{eq:s1adv}) then simplify considerably. We obtain
\begin{equation}
  \dot{\Phi_m}=                           
  \frac{128 s^3 \lambda e^{-s(\Lambda-\Delta)}}{(1+d)\mathcal{E}_m \Lambda} 
  \left(\begin{array}{cc}
    \left((1-d)s\Lambda\right)^{-1} & 0 \\
    0 & 1
  \end{array} \right) \Phi_m \,,
  \label{eq:Phieq2}
\end{equation}
and
\begin{equation}
  \sigma_1 = -r + \frac{r}{2} \left((1-d^2)-\frac{4}{s \Lambda} \right)
    \left( 1+d-\frac{2}{s \Lambda}\right)^{-1} \, .
\end{equation}

As a consequence, the critical value for friction is, at leading order,
\begin{equation}
  r_c = \frac{256 s^3 \lambda e^{-s(\Lambda-\Delta)}}{(1+d)^2 \Lambda} \, .
  \label{eq:rcadv}
\end{equation}
Notice that this relation is not valid for small $d$ and does not 
match (\ref{critr}) for $d=\Delta=0$.

\bibliography{../gfd}

\begin{thebibliography}{10}
\expandafter\ifx\csname url\endcsname\relax
  \def\url#1{\texttt{#1}}\fi
\expandafter\ifx\csname urlprefix\endcsname\relax\def\urlprefix{URL }\fi

\bibitem{Rhines:75}
P.~Rhines, Waves and turbulence on a beta-plane, J.~Fluid Mech. 69 (1975)
  417--443.

\bibitem{Williamson:78}
G.~Williamson, Planetary circulation \protect{I}. \protect{Barotropic}
  representation of \protect{Jovian} and terrestrial turbulence, J.~Atmos. Sci.
  35 (1978) 1399--1426.

\bibitem{Rhines:94}
P.~Rhines, Jets, Chaos 4 (1994) 313--339.

\bibitem{Vallis:93}
G.~Vallis, M.~Maltrud, Generation of mean flow and jets on a beta-plane and
  over topography, J.~Phys. Ocean. 23 (1993) 1346--1362.

\bibitem{Manfroi:99}
A.~Manfroi, W.~Young, Slow evolution of zonal jets on the beta-plane, J.~Atmos.
  Sci. 56 (1999) 784--800.

\bibitem{Frisch:96}
U.~Frisch, B.~Legras, B.~Villone, Large-scale \protect{Kolmogorov} flow on the
  beta-plane and resonant wave interactions, Physica D 94 (1996) 36--56.

\bibitem{Meshalkin:61}
L.~Meshalkin, Y.~Sinai, Investigation of the stability of a stationary solution
  of a system of equations for the plane movement of an incompressible viscous
  liquid, Appl. Math. Mech. 25.

\bibitem{Sivashinsky:85}
G.~Sivashinsky, Weak turbulence in periodic flows, Physica D 17 (1985)
  243--255.

\bibitem{Nepomnyashchy:76}
A.~A. Nepomnyashchyi, On the stability of the secondary flow of a viscous fluid
  in an infinite domain, Appl. Math. Mech. 40 (1976) 886--891.

\bibitem{She:87}
Z.~She, Metastability and vortex pairing in the \protect{Kolmogorov} flow,
  Phys. Lett. A 124 (1987) 161--164.

\bibitem{Legras:99}
B.~Legras, U.~Frisch, B.~Villone, Dispersive stabilization of the inverse
  cascade fo the \protect{Kolmogorov} flow, Phys. Rev. Lett. 82~(22) (1999)
  4440--4443.

\bibitem{Kawasaki:82}
K.~Kawasaki, T.~Ohta, Kink dynamics in one-dimensional nonlinear systems,
  Physica A 116 (1982) 573.

\bibitem{Pedlosky:87}
J.~Pedlosky, Geophysical Fluid Dynamics, 2nd Edition, Springer Verlag,
  New-York, 1987.

\bibitem{Manfroi:02}
A.~Manfroi, W.~Young, Stability of $\beta$-plane \protect{Kolmogorov} flow,
  Physica D 162~(3-4) (2002) 208--232.

\bibitem{Legras:02a}
B.~Legras, On the stability of $\beta$-plane \protect{Kolmogorov} flow,
  preprint.

\bibitem{Bates:95}
P.~Bates, J.~Xun, Metastable patterns for the \protect{Cahn-Hilliard} equation:
  protect{Part II}, J. Diff. Eq. 117 (1995) 165--216.

\bibitem{Frisch:95}
U.~Frisch, B.~Legras, B.~Villone, Large-scale dynamics of the
  \protect{Kolmogorov} flow on the beta-plane, in: R.~Benzi (Ed.), Advances in
  Turbulence, Proceedings Fifth European Turbulence Conference, Kluwer, 1995,
  pp. 138--140.

\bibitem{Kraichnan:67}
R.~H. Kraichnan, Inertial ranges in two-dimensional turbulence, Phys. Fluids 10
  (1967) 1417--1423.

\bibitem{Legras:88}
B.~Legras, P.~Santangelo, R.~Benzi, High-resolution numerical experiments for
  forced two-dimensional turbulence, Europhys. Lett. 5~(1) (1988) 37--42.

\end{thebibliography}

\setcounter{table}{0}
\renewcommand{\thetable}{\arabic{table}}
\newpage
\begin{longtable}{|@{}c@{}|@{}c@{}|@{}c@{}|@{}c@{}|@{}c@{}|
@{}c@{}|@{}c@{}|@{}c@{}|@{}c@{}|}
\caption{Critical values of friction and $\beta$ as a function
  of the number of unstable modes $n$ (or equivalently the size of the
  domain $L=n(3/2)^1/2$) and the wavenumber $N$ of the stationary
  solution. The columns $r_c^\mathrm{an}$ and $\beta_c^\mathrm{an}$
  show the analytical predictions given by (\ref{critr}) and
  (\ref{betac}) for $m=1$. The columns $r_c^\mathrm{pert}$ and
  $\beta_c^\mathrm{pert}$ show the critical values obtained by
  numerically solving the perturbation problem as indicated in
  Section~\ref{s:numpert}, again for $m=1$.  The columns
  $r_c^\mathrm{num}$ and $\beta_c^\mathrm{num}$ show the critical
  values obtained from the direct numerical stability study of the
  complete Cahn--Hilliard equation, as indicated in
  Section~\ref{s:numstab}.
   \label{t:fricbeta}} \\
\hline 
\( n \)&
\( N \)&
\( e^{-s\Lambda/2 } \)&
\( r_c^\mathrm{an} \)&
\( r_c^\mathrm{pert} \)&
\( r_c^\mathrm{num} \)&
\( \beta_c^\mathrm{an} \)&
\( \beta_c^\mathrm{pert} \)&
\( \beta_c^\mathrm{num} \)\\
\hline
\hline 
7&
2&
\( 4\,10^{-4} \)&
\( 6.8478\, 10^{-7} \)&
\( 6.8484\, 10^{-7} \)&
\( 6.848\, 10^{-7} \)&
\( 1.06\, 10^{-4} \)&
\( 2.0522\, 10^{-4} \)&
\( 2.0523\, 10^{-4} \)\\
\hline 
&
3&
\( 6\,10^{-3} \)&
\( 2.7469\, 10^{-4} \)&
\( 2.7705\, 10^{-4} \)&
\( 2.779\, 10^{-4} \)&
\( 5.88\, 10^{-3} \)&
\( 1.6329\, 10^{-2} \)&
\( 1.6385\, 10^{-2} \)\\
\hline 
&
4&
\( 2\,10^{-2} \)&
\( 5.5655\, 10^{-3} \)&
\( 5.9149\, 10^{-3} \)&
\( 6.147\, 10^{-7} \)&
\( 5.58\, 10^{-2} \)&
\( 0.216 \)&
\( 0.222 \)\\
\hline 
&
5&
\( 5\,10^{-2} \)&
\( 3.4908\, 10^{-2} \)&
\( 3.9614\, 10^{-2} \)&
\( 5.139\, 10^{-2} \)&
\( 0.260 \)&
\( 1.473 \)&
\( 1.379 \)\\
\hline 
20&
2&
\( 2\,10^{-10} \)&
\( 6.8832\, 10^{-20} \)&
\( 6.8825\, 10^{-20} \)&
NA&
\( 3.93\, 10^{-12} \)&
\( 4.82\, 10^{-22} \)&
NA\\
\hline 
&
3&
\( 4\,10^{-7} \)&
\( 4.1846\, 10^{-13} \)&
\( 4.1847\, 10^{-13} \)&
NA&
\( 2.50\, 10^{-8} \)&
\( 3.42\, 10^{-8} \)&
NA\\
\hline 
&
4&
\( 2\,10^{-5} \)&
\( 1.0438\, 10^{-9} \)&
\( 1.0439\, 10^{-9} \)&
\( 1.009\, 10^{-9} \)&
\( 2.41\, 10^{-6} \)&
\( 3.780\, 10^{-6} \)&
\( 3.716\, 10^{-6} \)\\
\hline 
&
5&
\( 10^{-4} \)&
\( 1.1756\, 10^{-7} \)&
\( 1.1756\, 10^{-7} \)&
\( 1.1751\, 10^{-7} \)&
\( 4.23\, 10^{-5} \)&
\( 7.782\, 10^{-5} \)&
\( 7.782\, 10^{-5} \)\\
\hline 
&
6&
\( 6\,10^{-4} \)&
\( 2.8135\, 10^{-6} \)&
\( 2.8143\, 10^{-6} \)&
\( 2.8144\, 10^{-6} \)&
\( 3.13\, 10^{-4} \)&
\( 6.858\, 10^{-4} \)&
\( 6.858\, 10^{-4} \)\\
\hline 
&
7&
\( 2\,10^{-3} \)&
\( 2.7738\, 10^{-5} \)&
\( 2.7778\, 10^{-5} \)&
\( 2.7795\, 10^{-5} \)&
\( 1.40\, 10^{-3} \)&
\( 3.706\, 10^{-2} \)&
\( 3.710\, 10^{-2} \)\\
\hline 
&
8&
\( 4\,10^{-3} \)&
\( 1.5681\, 10^{-4} \)&
\( 1.5750\, 10^{-4} \)&
\( 1.580\, 10^{-4} \)&
\( 4.52\, 10^{-3} \)&
\( 1.468\, 10^{-2} \)&
\( 1.479\, 10^{-2} \)\\
\hline 
&
9&
\( 7\,10^{-3} \)&
\( 6.1103\, 10^{-4} \)&
\( 6.1582\, 10^{-4} \)&
\( 6.242\, 10^{-4} \)&
\( 1.18\, 10^{-2} \)&
\( 4.676\, 10^{-2} \)&
\( 4.827\, 10^{-2} \)\\
\hline 
&
10&
\( 10^{-2} \)&
\( 1.8329\, 10^{-3} \)&
\( 1.8416\, 10^{-3} \)&
\( 1.921\, 10^{-3} \)&
\( 2.68\, 10^{-2} \)&
\( 0.124 \)&
\( 0.138 \)\\
\hline 
&
11&
\( 2\,10^{-2} \)&
\( 4.5418\, 10^{-3} \)&
\( 4.4508\, 10^{-3} \)&
\( 4.965\, 10^{-3} \)&
\( 5.24\, 10^{-2} \)&
\( 0.287 \)&
\( 0.373 \)\\
\hline 
&
12&
\( 2\,10^{-2} \)&
\( 9.7452\, 10^{-3} \)&
\( 8.9096\, 10^{-3} \)&
\( 1.135\, 10^{-2} \)&
\( 9.58\, 10^{-2} \)&
\( 0.551 \)&
\( 1.01 \)\\
\hline 
&
13&
\( 3\,10^{-2} \)&
\( 1.8708\, 10^{-2} \)&
\( 1.4927\, 10^{-2} \)&
\( 2.392\, 10^{-2} \)&
\( 0.16 \)4&
\( 0.951 \)&
\( 2.11 \)\\
\hline
\end{longtable}


\newpage
\begin{longtable}{|c|c|c|c|c|c|c|c|}
\caption{Values of the phase speed of the basic solution and the frequency 
  of the $m=1$ mode as a function of $n$ and $N$. The columns
  $c1^\mathrm{an}/\beta$ and $\mu_1^\mathrm{an}/\beta$ show the
  analytical predictions given by (\ref{speed}) and (\ref{m1a}).  The
  columns $c_1^\mathrm{pert}/\beta$ and $\mu_1^\mathrm{pert}/\beta$
  show the values obtained by numerically solving the eigenvalue
  problem.  $\mu_1^\mathrm{an}$ and $\mu_1^\mathrm{pert}$ are
  calculated for $m=1$.  The columns $c^\mathrm{num}/\beta$ and
  $\mu^\mathrm{num}/\beta$ show the values obtained from the direct
  numerical stability study of the complete Cahn--Hilliard equation
  near the critical value of $\beta$. $\mu^\mathrm{num}$ is the
  frequency of the critical mode.
   \label{t:cmu}} \\
\hline
\( n \)&
\( N \)&
\( c_1^\mathrm{an}/\beta \)&
\( c_1^\mathrm{pert}/\beta \)&
\( c^\mathrm{num}/\beta \)&
\( \mu_1^\mathrm{an}/\beta \)&
\( \mu_1^\mathrm{pert}/\beta \)&
\( \mu^\mathrm{num}/\beta \)\\
\hline
\hline 
7&
2&
-17.6492&
-17.6492&
-17.649&
-6.01296&
-6.01298&
-6.0127\\
\hline 
&
3&
-8.05199&
-8.0485&
-8.0486&
-2.21915&
-2.22148&
-2.222\\
\hline 
&
4&
-4.61026&
-4.57267&
-4.5730&
-1.16117&
-1.18408&
-1.1876\\
\hline 
&
5&
-3.11049&
-2.93650&
-2.9368&
-0.63701&
-0.73943&
-0.74685\\
\hline 
20&
2&
-133.043&
-133.043&
-133.04&
-16.4572&
-16.4571&
NA\\
\hline 
&
3&
-60.8209&
-60.8209&
-60.821&
-5.86179&
-5.86179&
NA\\
\hline 
&
4&
-35.0024&
-35.0024&
-35.002&
-3.08063&
-3.08063&
-3.0806\\
\hline 
&
5&
-22.8122&
-22.8122&
-22.812&
-1.92726&
-1.92726&
-1.9273\\
\hline 
&
6&
-16.0670&
-16.0669&
-16.067&
-1.33119&
-1.33123&
-1.3312\\
\hline 
&
7&
-11.9316&
-11.9310&
-11.931&
-0.979702&
-0.979894&
-0.97994\\
\hline 
&
8&
-9.20895&
-9.20716&
-9.2074&
-0.753012&
-0.753716&
-0.75386\\
\hline 
&
9&
-7.32184&
-7.31636&
-7.3170&
-0.596714&
-0.59863&
-0.59900\\
\hline 
&
10&
-5.96367&
-5.95021&
-5.9514&
-0.482911&
-0.487182&
-0.48791\\
\hline 
&
11&
-4.95971&
-4.93139&
-4.9335&
-0.395862&
-0.404176&
-0.40547\\
\hline 
&
12&
-4.20511&
-4.15177&
-5.1550&
-0.325842&
-0.340601&
-0.34263\\
\hline 
&
13&
-3.63500&
-3.54226&
-3.5452&
-0.266160&
-0.290808&
-0.2927\\
\hline
\end{longtable}

\newpage
\setlongtables
\begin{longtable}{|c|c|c|c|c|}
\caption{Distribution of solutions as a function of $N$ for $n=20$, 
   $\beta=0$ and increasing values of $r$. The statistics is calculated out of
   an ensemble of 100 cases for each value of $r$ and $A$. The cases are
   obtained by varying the seed
   of the random generator. The fourth and fifth columns show the percentages
   of solutions ending on the stable $N$-pair solutions with $N$ 
   in the third column. The first column gives the critical linear
   stability $r_c$ as function of $N$, provided by the numerical solution
   to the perturbative problem (see Section~\ref{s:persta}).
   \label{t:r}} \\
\hline
\( r_{c} \)&
\( r \)&
\( N \)&
\( A=0.1 \)&
\( A=1. \)\\
\hline
\hline 
\( 2.81\, 10^{-6} \)&
\multicolumn{1}{c}{}&
\multicolumn{1}{c}{6}&
\multicolumn{1}{c}{}&
\multicolumn{1}{c|}{}\\
\hline 
&
\( 10^{-5} \)&
4&
4\%&
19\%\\
\hline 
&
&
5&
81\%&
71\%\\
\hline 
&
&
6&
15\%&
10\%\\
\hline 
&
\( 1.78\, 10^{-5} \)&
4&
1\%&
8\%\\
\hline 
&
&
5&
57\%&
69\%\\
\hline 
&
&
6&
42\%&
23\%\\
\hline 
\( 2.78\, 10^{-5} \)&
\multicolumn{1}{c}{}&
\multicolumn{1}{c}{7}&
\multicolumn{1}{c}{}&
\multicolumn{1}{c|}{}\\
\hline 
&
\( 3.16\, 10^{-5} \)&
4&
&
1\%\\
\hline 
&
&
5&
44\%&
52\%\\
\hline 
&
&
6&
56\%&
41\%\\
\hline 
&
\( 5.62\, 10^{-5} \)&
5&
23\%&
30\%\\
\hline 
&
&
6&
74\%&
67\%\\
\hline 
&
&
7&
3\%&
3\%\\
\hline 
&
\( 10^{-4} \)&
5&
9\%&
16\%\\
\hline 
&
&
6&
76\%&
75\%\\
\hline 
&
&
7&
15\%&
9\%\\
\hline 
\( 1.57\, 10^{-4} \)&
\multicolumn{1}{c}{}&
\multicolumn{1}{c}{8}&
\multicolumn{1}{c}{}&
\multicolumn{1}{c|}{}\\
\hline 
&
\( 1.78\, 10^{-4} \)&
5&
1\%&
2\%\\
\hline 
&
&
6&
50\%&
65\%\\
\hline 
&
&
7&
49\%&
33\%\\
\hline 
&
\( 3.16\, 10^{-4} \)&
6&
18\%&
34\%\\
\hline 
&
&
7&
75\%&
64\%\\
\hline 
&
&
8&
7\%&
2\%\\
\hline 
&
\( 5.62\, 10^{-4} \)&
6&
1\%&
19\%\\
\hline 
&
&
7&
63\%&
62\%\\
\hline 
&
&
8&
36\%&
2\%\\
\hline 
\( 6.16\, 10^{-4} \)&
\multicolumn{1}{c}{}&
\multicolumn{1}{c}{9}&
\multicolumn{1}{c}{}&
\multicolumn{1}{c|}{}\\
\hline 
&
\( 10^{-3} \)&
6&
&
6\%\\
\hline 
&
&
7&
29\%&
40\%\\
\hline 
&
&
8&
69\%&
54\%\\
\hline 
&
&
9&
2\%&
\\
\hline 
&
\( 1.78\, 10^{-3} \)&
7&
1\%&
20\%\\
\hline 
&
&
8&
74\%&
71\%\\
\hline 
&
&
9&
25\%&
9\%\\
\hline 
\( 1.84\, 10^{-3} \)&
\multicolumn{1}{c}{}&
\multicolumn{1}{c}{10}&
\multicolumn{1}{c}{}&
\multicolumn{1}{c|}{}\\
\hline 
&
\( 3.16\, 10^{-3} \)&
7&
&
3\%\\
\hline 
&
&
8&
26\%&
57\%\\
\hline 
&
&
9&
68\%&
69\%\\
\hline 
&
&
10&
6\%&
\\
\hline 
\( 4.45\, 10^{-3} \)&
\multicolumn{1}{c}{}&
\multicolumn{1}{c}{11}&
\multicolumn{1}{c}{}&
\multicolumn{1}{c|}{}\\
\hline 
&
\( 5.62\, 10^{-3} \)&
8&
2\%&
20\%\\
\hline 
&
&
9&
66\%&
70\%\\
\hline 
&
&
10&
32\%&
10\%\\
\hline 
\( 8.90\, 10^{-3} \)&
\multicolumn{1}{c}{}&
\multicolumn{1}{c}{12}&
\multicolumn{1}{c}{}&
\multicolumn{1}{c|}{}\\
\hline 
&
\( 10^{-2} \)&
8&
&
3\%\\
\hline 
&
&
9&
20\%&
51\%\\
\hline 
&
&
10&
66\%&
45\%\\
\hline 
&
&
11&
14\%&
1\%\\
\hline 
\( 1.49\, 10^{-2} \)&
\multicolumn{1}{c}{}&
\multicolumn{1}{c}{13}&
\multicolumn{1}{c}{}&
\multicolumn{1}{c|}{}\\
\hline 
&
\( 1.78\, 10^{-2} \)&
9&
1\%&
9\%\\
\hline 
&
&
10&
33\%&
59\%\\
\hline 
&
&
11&
62\%&
32\%\\
\hline 
&
&
12&
4\%&
\\
\hline 
&
\( 3.16\, 10^{-2} \)&
10&
3\%&
17\%\\
\hline 
&
&
11&
48\%&
59\%\\
\hline 
&
&
12&
48\%&
23\%\\
\hline 
&
&
13&
1\%&
1\%\\
\hline
\end{longtable}

\newpage
\begin{longtable}{|c|c|c|c|c|}
\caption{Same as Table~\ref{t:beta} but for the 
   distribution of solutions as a function of $N$ for $n=20$, 
   $r=0$ and increasing values of $\beta$.
\label{t:beta}} \\
\hline 
beta\_c\(  \)&
beta\( \beta  \)&
\( N \)&
\( A=0.1 \)&
\( A=1. \)\\
\hline
\hline 
\( 7.78\, 10^{-5} \)&
\multicolumn{1}{c}{}&
\multicolumn{1}{c}{5}&
\multicolumn{1}{c}{}&
\multicolumn{1}{c|}{}\\
\hline 
&
\( 10^{-4} \)&
3&
12\%&
13\%\\
\hline 
&
&
4&
85\%&
87\%\\
\hline 
&
&
5&
3\%&
\\
\hline 
\( 6.86\, 10^{-4} \)&
\multicolumn{1}{c}{}&
\multicolumn{1}{c}{6}&
\multicolumn{1}{c}{}&
\multicolumn{1}{c|}{}\\
\hline 
&
\( 10^{-3} \)&
4&
17\%&
36\%\\
\hline 
&
&
5&
82\%&
63\%\\
\hline 
&
&
6&
1\%&
1\%\\
\hline 
\( 3.71\, 10^{-3} \)&
\multicolumn{1}{c}{}&
\multicolumn{1}{c}{7}&
\multicolumn{1}{c}{}&
\multicolumn{1}{c|}{}\\
\hline 
&
\( 10^{-2} \)&
5&
&
2\%\\
\hline 
&
&
6&
74\%&
68\%\\
\hline 
&
&
7&
26\%&
30\%\\
\hline 
\( 1.47\, 10^{-2} \)&
\multicolumn{1}{c}{}&
\multicolumn{1}{c}{8}&
\multicolumn{1}{c}{}&
\multicolumn{1}{c|}{}\\
\hline 
\( 4.68\, 10^{-2} \)&
\multicolumn{1}{c}{}&
\multicolumn{1}{c}{9}&
\multicolumn{1}{c}{}&
\multicolumn{1}{c|}{}\\
\hline 
&
\( 0.1 \)&
7&
2\%&
10\%\\
\hline 
&
&
8&
56\%&
78\%\\
\hline 
&
&
9&
42\%&
12\%\\
\hline 
\( 0.124 \)&
\multicolumn{1}{c}{}&
\multicolumn{1}{c}{10}&
\multicolumn{1}{c}{}&
\multicolumn{1}{c|}{}\\
\hline 
\( 0.288 \)&
\multicolumn{1}{c}{}&
\multicolumn{1}{c}{11}&
\multicolumn{1}{c}{}&
\multicolumn{1}{c|}{}\\
\hline 
\( 0.551 \)&
\multicolumn{1}{c}{}&
\multicolumn{1}{c}{12}&
\multicolumn{1}{c}{}&
\multicolumn{1}{c|}{}\\
\hline 
&
\( 1. \)&
7&
&
1\%\\
\hline 
&
&
8&
&
2\%\\
\hline 
&
&
9&
&
17\%\\
\hline 
&
&
10&
16\%&
26\%\\
\hline 
&
&
11&
65\%&
46\%\\
\hline 
&
&
12&
18\%&
8\%\\
\hline 
&
&
13&
1\%&
\\
\hline
\end{longtable}

\end{document}